\documentclass[LaM,binding=0.6cm,oneside, noexaminfo, english]{sapthesis}

\usepackage{microtype}
\usepackage[thinlines]{easytable}
\usepackage{verbatim}

\usepackage{hyperref}
\hypersetup{pdftitle={Usage example of the Sapthesis class for a Laurea Magistrale thesis in English},pdfauthor={Andrea Moscatelli}}

\usepackage{lipsum}
\usepackage{curve2e}
\usepackage{mathrsfs}

\definecolor{gray}{gray}{0.4}

\usepackage{float}
\usepackage{algorithm}
\usepackage[noend]{algpseudocode}

\usepackage{tabu}
\usepackage{todonotes}
\usepackage{multirow}
\usepackage{graphicx}
\graphicspath{ {./Images/} }

\usepackage{blindtext}

\title{Modeling of negative protein-protein \\ interactions: methods and experiments}
\author{Andrea Moscatelli}
\IDnumber{1667647}
\course{Computer Science}
\courseorganizer{Facolt\`a di Ingegneria dell'Informazione, Informatica e Statistica}
\AcademicYear{2018/2019}
\copyyear{2019}
\advisor{Prof. Paola Velardi}
\coadvisor{Prof. Giovanni Stilo}
\authoremail{andrea.moscatelli95@gmail.com}

\versiondate{\today}

\begin{document}
\newcommand{\algorithmicoutput}{    \textbf{Output}}
\newcommand{\OUTPUT}{\item[\algorithmicoutput]}
\newcommand{\algorithmicinput}{\tab \textbf{Input}}
\newcommand{\INPUT}{\item[\algorithmicinput]}
\newcommand{\algorithmicinit}{}
\newcommand{\INIT}{\item[\algorithmicinit]}
\frontmatter

\maketitle

\begin{acknowledgments}
    First, I would like to thank my thesis advisor, Professor Paola Velardi, who, with her integrity and transparency, allowed me to grow humanly and professionally. Her feedback, advice and suggestions were priceless.\\
    I would also like to thank Professor Yang Yu-Liu and Professor Song (Stephen) Yi. This work would not be as successful without their participation and contribution.\\
    Moreover, I express my profound gratitude to my family, which unconditionally supported me during my whole life, allowing me to achieve \textbf{our} results. The best is yet to come.\\
    Finally, I would also like to thank all my friends and colleagues, which, in various ways, have helped me while I was completing my thesis. Thanks to Luca, which shared with me joys, sorrows, tv series and video games. Thanks to Monti, 3 Etti and Alessio, which backed me all the way. Thanks to Motino, which was always there for me. And a big thanks also to all the others: Lorenzo, Irene, Alessio, Maurizio, Perna, Martina, Mary, Giulia, Rubia, Macchietta, Flavio, Roberto, Antonio, Federica, Giuseppe, Vicky, Marianna, Silvia, Omar, Ilaria, Jacopo. 

\end{acknowledgments}

\tableofcontents

\mainmatter

\chapter{Introduction}
\label{introduction_chapter}

The interactions between proteins, also referred as protein-protein interactions (PPIs), are of fundamental importance for the human body, and the knowledge about their existence could provide useful insights when performing critical tasks, as drug target developing and therapy design.

However, the high-throughput laboratory experiments generally used to discover new protein-protein interactions are very costly and time consuming, stressing the need of new computational methods able to predict high-quality PPIs. These methods have to face two main problems: (i) the very low number of PPIs already known and (ii) the lack of high-quality negative protein-protein interactions (i.e. proteins that are known to not interact). The former is due to the high number of PPIs in the human body and the high cost of the PPIs detection laboratory experiments. Instead, the latter is usually overlooked by the PPIs prediction systems, causing a significant bias in the performances and metrics.

This work is particularly focused on the issue regarding the absence of negative interactions. This problem is of crucial importance since, to predict high-quality PPIs, the computational systems have to learn from both positive and negative instances, and, even when only positive examples are exploited in training, negative data are still needed to evaluate performances in an appropriate manner. 

Our results show that some methods for generating reliable negative instances are more effective than others and that the performances reported by the PPIs prediction systems in literature are usually overestimated, mainly because of the negative interactions used in the training and testing phases.

The thesis is organized as follows. Chapter \ref{chapter:backgrounds} contains an overview of the computer science and biological backgrounds needed to properly understand this work. Chapter \ref{chapter:datasets} describes the datasets used, after a brief and general introduction on the PPIs datasets. Chapter \ref{chapter:ppi_prediction_review} introduces the PPIs prediction task, describing also the computational methods implemented to solve it and the main issues that these methods have to tackle. Chapter \ref{chapter:negative_interactions_problem} presents the negative interactions issue, remarking its importance and describing in detail the methods generally used to generate negative instances, along with their strengths and weaknesses. Chapter \ref{chapter:negative_methods_comparison} contains a comparison between two of the methods described in Chapter \ref{chapter:negative_interactions_problem}, analyzing their performances when very reliable datasets are used. In Chapter \ref{chapter:ppi_prediction_model}, several experiments are presented. First, different features are described, also analysing their importance with respect to the PPIs prediction task. Then, a PPIs prediction system is presented and is tested on reliable validation sets, achieving good accuracy. Finally, two of the state-of-the-art PPIs prediction systems are tested, showing how their performances drop when reliable negative validation sets are used. 
Chapter \ref{chapter:textmining_survey} contains a survey on protein-protein interactions extraction from biomedical literature, with a particular focus on the extraction of negative PPIs. Finally, Chapter \ref{chapter:conclusions} contains the conclusions and the future directions.

\chapter{Backgrounds needed}
\label{chapter:backgrounds}

The following two sections contain a brief overview of the backgrounds needed to understand this work.

\section{Computer Science background}
\label{section:cs_background}

A \textbf{Graph} is a mathematical structure that is often used to define a set of objects and their relations. A graph consists of:

\begin{enumerate}
    \item A finite set of points, called \textbf{nodes} or \textbf{vertices}.

    \item A finite set of lines, called \textbf{edges} or \textbf{arcs}. One important property of the edges is that they can be both \textit{directed} and \textit{undirected}. If an edge (A, B) is \textit{undirected}, it means that A is linked to B and B is linked to A. Instead, if the edge (A, B) is \textit{directed}, it only means that there is a link from A to B. 
    
    Another important property of the edges is the weight. In a binary graph, an edge can only take two values: 0 (the edge does not exist) or 1 (the edge exists). In a weighted graph, instead, a numerical value is assigned to each edge. Finally, a graph can also be signed, meaning that a sign is attached to each edge (allowing negative edges). 
    
\end{enumerate}

\begin{figure}[]

    \centering
    \includegraphics[width=0.90\textwidth]{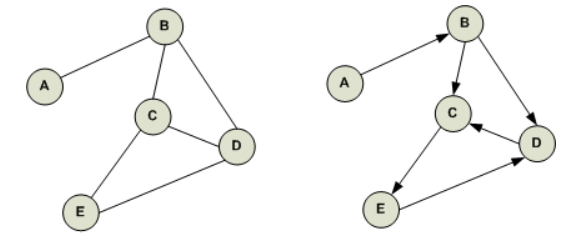}
    \caption{The difference between two graphs based on their directionality. The figure on the left shows an undirected graph, while the figure on the right shows a directed graph\protect\footnotemark.}
    \label{fig:graph_directed_vs_undirected}
\end{figure}

\footnotetext{\url{https://codingwithalex.com/introduction-to-graphs/}}

One of the most important properties of a graph is its density, which is represented as the ratio between the number of edges and the maximum possible number of edges.

Formally, for undirected graphs: \\

\begin{equation}
\label{undirected_graph_density_equation}
    \frac{2\left | E \right |}{\left | V \right | \left ( \left | V \right | - 1\right )}
\end{equation}
\\

Where $\left | E \right |$ is the number of edges and $\left | V \right |$ is the number of nodes.
\section{Biological background}
\label{section:biological_background}

Proteins are complex molecules that perform several functions in the body\footnote{\url{https://ghr.nlm.nih.gov/primer/howgeneswork/protein}}. Among other things, they are responsible for DNA replication, molecule transportation and for the structure definition of the cells and the organisms. 

A protein rarely acts alone, since it tends to establish physical contacts, called \textbf{interactions}, with other proteins in order to accomplish its functions. Indeed, it has been shown that more than 80\% of the proteins act in complexes, rather than alone \cite{methods_analysis}. 

These interactions are of crucial importance for several reasons\footnote{\url{https://www.ebi.ac.uk/training/online/course/protein-interactions-and-their-importance/protein-protein-interactions/importance-molecular-i}}:

\begin{enumerate}
    \item They help to understand a protein's function and behaviour.
    
    \item They help to identify the unknown characteristics of a specific protein, based on its interactions. For example, the function of a protein could be predicted on the basis of the proteins interacting with it.
    
    \item They represent the edges of the Protein-protein Interaction Networks, which are used to tackle different problems, from drug design to protein-protein interactions prediction.
    
\end{enumerate}

From a structural point of view, the proteins are composed of amino acids, that can be defined as organic molecules consisting of a basic amino group (-NH2), an acidic carboxyl group (-COOH), and an organic R group (or side chain)\footnote{\url{https://www.britannica.com/science/amino-acid}}. Since they contain the information about the protein's structure, the amino acids are often used in several tasks (e.g. protein-protein interactions prediction). In total, 20 different amino acids are considered the essential building blocks of all proteins.

Another important concept that must be mentioned is the \textit{protein domain}, that is a conserved region of a given protein sequence that can evolve, function, and exist independently of the rest of the protein chain\footnote{\url{https://en.wikipedia.org/wiki/Protein_domain}}. The domains have a length that varies from 50 to 250 amino acids, and are generally responsible for a specific function or interaction. Furthermore, the domains are very important for the protein-protein interactions prediction task, since only proteins with complementary domains can interact. Moreover, as shown by Wang et al. \cite{domain_insights_diseases}, domains also provide insights into human genetic disease. 

\begin{figure}[]
    \centering
    \includegraphics[width=0.40\textwidth]{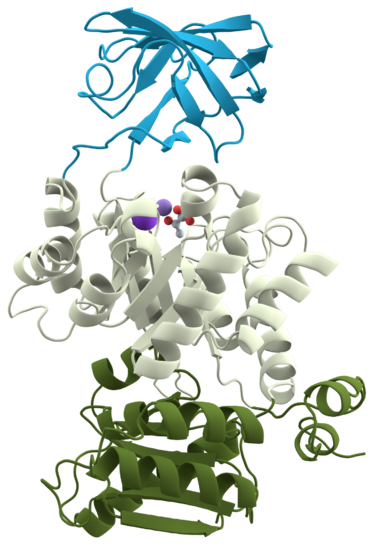}
    \caption{The protein Pyruvate Kinase, which has three different domains: an all-$\beta$ nucleotide binding domain (in blue), an $\alpha / \beta$-substrate binding domain (in grey) and an $\alpha / \beta$-regulatory domain (in green)\protect\footnotemark.}
    \label{fig:pyruvate_kinase_protein_domains}
\end{figure}

A visual example of the protein domains is shown in Figure \ref{fig:pyruvate_kinase_protein_domains}, where is represented the Pyruvate Kinase, a protein containing three different domains.

The knowledge of graphs can be combined with the knowledge of proteins to create a Protein-protein Interaction Network (PIN), in which the nodes are the proteins and there is an edge between two proteins A and B (formally \textit{(A, B)}) when they are known to physically interact.

\footnotetext{\url{https://en.wikipedia.org/wiki/Protein_domain}}

Generally, in a PIN, there can be two possible types of edges:

\begin{enumerate}
    \item Positive edges, between two proteins that are known to physically interact.
    \item Negative edges, between two proteins that are known to \textbf{not} interact.
\end{enumerate}

Due to their high importance, the protein-protein interactions are constantly studied and experimentally detected. One of the most popular and effective methods for detecting protein-protein interactions is the Yeast two-Hybrid (Y2H) screen.

The Y2H relies on the expression of a specific reporter gene\footnote{\url{https://www.singerinstruments.com/application/yeast-2-hybrid/}}. This gene can be activated by the binding of a transcription factor, composed of two independent domains: the DNA-binding domain (DB) and the activation domain (AD). These domains, since they are functionally and structurally independent, can be fused to two different proteins. One protein, referred as \textit{bait}, is fused to the DB domain, whereas the other, referred as \textit{prey}, is fused to the AD.

\begin{figure}[]
    \centering
    \includegraphics[width=0.70\textwidth]{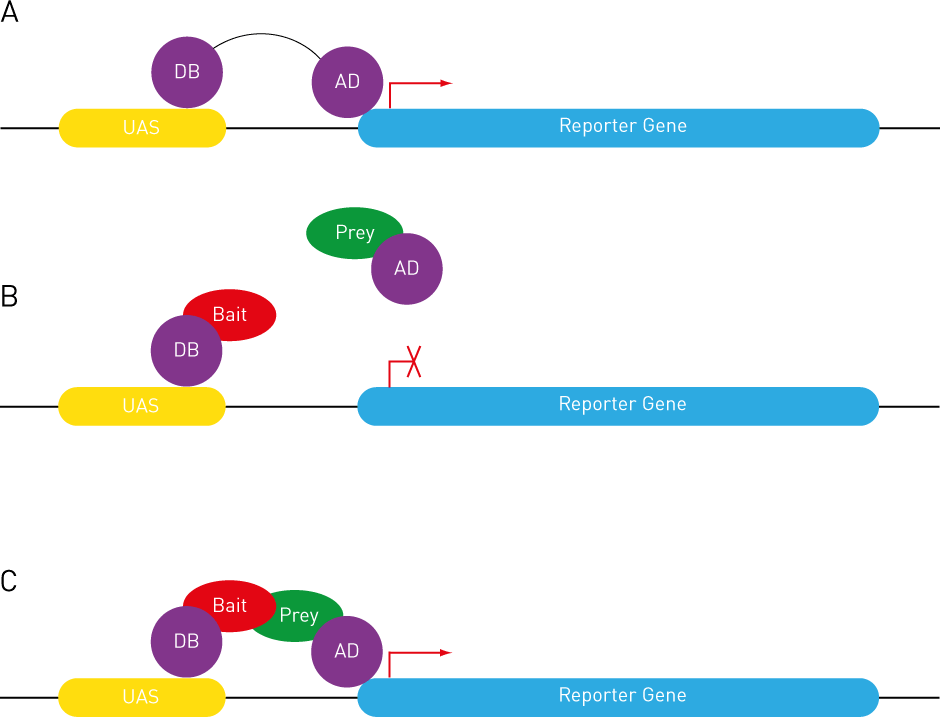}
    \caption{An overview of the Yeast two-Hybrid assay\protect\footnotemark. The bait (in red) is fused to the DNA-binding domain whereas the prey (in green) is fused to the Activation domain. If these proteins do not interact, the expression of the reporter gene is not activated (B), while, if they interact, the reporter gene expression is activated by the activation domain (C).}
    \label{fig:y2h_assay}
\end{figure}

\footnotetext{\url{https://www.singerinstruments.com/application/yeast-2-hybrid/}}

As can be seen from figure \ref{fig:y2h_assay}, only when the bait and prey interact the expression of the reporter gene is possible. Instead, if the two proteins do not interact, the the activation domain is not able to localize the reporter gene to drive gene expression.
The Y2H experiments can be performed also in a large-scale way, in which a single bait can be tested, for example, against an array of preys \cite{largescale_y2h}, leading to the discovery of several interaction partners.

\chapter{Datasets}
\label{chapter:datasets}

One of the problems encountered when retrieving the state-of-the-art systems for PPI link prediction is that often different datasets were used among different systems, making a fair comparison impossible. 

Indeed, there are a large number of protein-protein interactions databases \cite{dbsurvey}. The main ones, here referred as \textit{primary databases}\footnote{Notice that some of them are old or slowly updated \cite{hprd, dip}}, obtain all of their data by curating peer-reviewed publications \cite{intact, hprd, biogrid, dip}. An example of primary database is the Biological General Repository for Interaction Datasets (BioGRID), which currently contains over 380 thousand unique protein-protein interactions only for the Homo Sapiens organism.

However, in each primary database there are interactions not contained in other primary databases. Therefore, there are other databases, here referred as \textit{secondary databases}, that integrate the contents of multiple primary databases \cite{hippie, string}. One of the most used secondary database is STRING \cite{string}, which contains also information about computationally predicted interactions. 

However, with high probability, some of the interactions contained in these databases (both primary and secondary) are false positives (i.e. they do not exist although they are reported), making the resulting databases highly biased. This is also shown in \cite{hi14}, where the authors conducted an experiment to measure the reliability of the interactions derived from medical literature. Specifically, they extracted 33,000 literature binary protein interactions and they divided them into those reported only in a single publication and detected by only a single method (formally LIT-BS), which accounted for two thirds of the total number of interactions, and those supported by multiple pieces of evidence (formally LIT-BM). Then, they used two protein-protein interaction detection methods, the mammalian protein-protein interaction trap \cite{mappit} and yeast two-hybrid assay \cite{y2h}, to test the reliability of these two sets (LIT-BS and LIT-BM). As a result, they discovered that the recovery rate for the pairs in LIT-BS was only a bit higher than the one of the random selected protein pairs used as negative control and considerably lower than the recovery rate of the pairs in LIT-BM, showing that only the interactions with multiple pieces of evidence should be considered reliable. At the time the paper was written, the number of these interactions (i.e. those having multiple pieces of evidence) was 11045, which is a very small number with respect to the the estimated size of the human PIN \cite{estimated_size_interactome}.

The proteins in the interactions contained in each database are always associated with an ID that univocally identify them. The most used protein IDs are the UniprotKB IDs, altough other identifiers, such as the Gene names and the Entrez Gene IDs are widely used. Therefore, when two databases using different protein identifiers are used, a mapping from the identifiers used by the first to the identifiers used by the other is required. For example, supposing that one database uses the UniprotKB identifiers while the other uses the Gene names, a specific protein would be identified as P31946 (UniprotKB ID) in the first database while it would be identified as YWHAB (gene name) in the second database.  

However, the mapping process sometimes could be a source of errors, due to the different structures of the identifiers databases. For example, a specific uniprotKB ID could not be mapped to any (or, more frequently, is mapped to more than one) gene name.

During this thesis, several datasets were used: HuRI, LIT-BM, HQND, and some other negative datasets generated by different methods, as will be discussed in Chapter \ref{chapter:negative_methods_comparison}. These datasets were used because, although smaller than the majority of primary and secondary databases, they are characterized by a high level of reliability.

\section{HuRI}
\label{section:huri_dataset}

HuRI is the Human Reference Interactome\footnote{http://interactome.baderlab.org/}. This dataset currently has three versions:

\begin{enumerate}
    \item \textbf{HI-I-05} \cite{hi05}(HI-05), which contains \textasciitilde 2800 interactions among \textasciitilde 1500 different proteins. 
    \item \textbf{HI-II-14} \cite{hi14}(HI-14), which contains \textasciitilde 14000 interactions among \textasciitilde 4500 different proteins.
    \item \textbf{HI-III-19} \cite{huri3}(HI-19), which contains \textasciitilde 54000 interactions among \textasciitilde 8400 different proteins.
\end{enumerate}

Although they are different versions of the same dataset, the overlap between their interactions and proteins is not so consistent, as shown in Tables \ref{table:proteins_overlap_table} and \ref{table:interactions_overlap_table}.

Indeed, regarding the proteins, only 79\% of the proteins in HI-05 are in HI-14, and same percentage is met when comparing the proteins in HI-14 with those in HI-19. Instead, regarding the interactions, only 27\% of the interactions of HI-05 are also in HI-14, while only 37\% of the interactions of HI-14 are contained in HI-19. \\

\begin{table}[]
\centering
\begin{tabular}{|c|c|c|c|}
\hline
Proteins & HI-05        & HI-14         & HI-19        \\ \hline
HI-05    & 1570 \textbf{(100\%)} & 1245 \textbf{(79\%)}  & 1259 \textbf{(80\%)}  \\
HI-14    & 1245 \textbf{(27\%)}  & 4523 \textbf{(100\%)} & 3584 \textbf{(79\%) } \\
HI-19    & 1259 \textbf{(14\%)}  & 3584 \textbf{(42\%)}  & 8490 \textbf{(100\%)} \\ \hline
\end{tabular}
\caption{Overlap among the proteins in the three versions of the HuRI dataset. The overlap between the proteins is more consistent with respect to the one of the interactions. Indeed, the 79\% of the proteins in HI-05 are contained in HI-14, and the same percentage is met when comparing the proteins in HI-14 with those in HI-19.}
\label{table:proteins_overlap_table}
\end{table}

\begin{table}[]
\centering
\begin{tabular}{|c|c|c|c|}
\hline
Interactions & HI-05        & HI-14         & HI-19        \\ \hline
HI-05    & 2770 \textbf{(100\%)} & 746 \textbf{(27\%)}  & 773 \textbf{(28\%)}  \\
HI-14    & 746 \textbf{(5\%)}  & 14614 \textbf{(100\%)} & 5393 \textbf{(37\%) } \\
HI-19    & 773 \textbf{(1.4\%)}  & 5393 \textbf{(10\%)}  & 54496 \textbf{(100\%)} \\ \hline
\end{tabular}
\caption{Overlap among the interactions in the three versions of the HuRI dataset. Although they represent different editions of the same dataset, only the 27\% of the interactions in HI-05 are contained in HI-14, whereas only the 37\% of the interactions contained in HI-14 are in HI-19.}
\label{table:interactions_overlap_table}
\end{table}

The last version of HuRI (released on April 2019), HI-III-19 (or HI-19), currently contains about 54496 interactions between 8490 different proteins \cite{huri3}. All the interactions in this dataset have been retrieved by a systematic yeast two-hybrid screening pipeline and have at least one piece of experimental data. Then, these interactions were further validated in multiple orthogonal assays, and their quality turned out to be comparable or greater than the quality of a set of interactions with more than two pieces of experimental evidence (LIT-BM).

In this dataset, the entries are often protein isoforms (i.e. protein variants originated from genetic differences), as Q15038-1. However, for some experiments, we had to convert each protein isoform to the original protein (e.g. Q15038-1 $ \rightarrow $ Q15038). 

The HuRI dataset also contains a wide range of useful information for each interaction, such as the biological roles of the two proteins in the experiment (e.g. \textit{bait}) and the confidence score of each interaction.

The proteins in this database are identified by their UniprotKB IDs and Gene names.

\section{LIT-BM}
\label{section:lit-bm}

LIT-BM is a database of protein-protein interactions that were retrieved from medical literature and supported by multiple pieces of evidence, of which at least one comes from a binary assay type \cite{hi14}. As previously explained, as a result of the experiment performed in \cite{hi14}, these interactions are the only ones that should be considered reliable among all the interactions extracted from medical literature. Indeed, this dataset has \textbf{comparable high quality} to that of HI-19. 

The LIT-BM dataset was downloaded from the Human Reference Protein Interactome website\footnote{http://interactome.baderlab.org/download} and contains 13030 protein interactions among \textasciitilde6000 proteins. The main difference between this dataset and the HuRI dataset is that this one, since the interactions in it are retrieved from the literature, is biased towards well studied proteins and contains more information in the dense zone of the human Protein Interaction Network, while the latter, being the result of a screening pipeline, is more systematic and unbiased. 

Among the information for each interaction in LIT-BM, is also present the mentha-score, that is a confidence score that takes into account all the experimental evidence retrieved from the different databases\footnote{http://mentha.uniroma2.it/browser/score.html}. 

The proteins in this database are identified by their UniprotKB IDs and Gene names.

\section{HQND}
\label{section:HQND}

While the two previous databases (HI-19 and LIT-BM) contain only positive interactions, the High Quality Negative Dataset contains only \textbf{negative} interactions, i.e. interactions that are known to not exist.

The negative interactions in this dataset are \textit{i)} not reported in the medical literature and \textit{ii)}  supported by at least 3 independent orthogonal experimental methods being negative. 

This dataset is not public and has been released to us by the University of Texas at Austin, in order to perform several experiments. Although it is very small (676 interactions among \textasciitilde 1200 proteins), its rarity arises from the fact that the interactions inside it are negative. Indeed, while there are plenty of datasets containing positive interactions, there are no \textbf{reliable} datasets containing negative interactions, as will be explained in detail in chapter \ref{chapter:negative_interactions_problem}. 

The proteins in this database are identified by their Gene IDs.

\chapter{PPIs prediction review in literature}
\label{chapter:ppi_prediction_review}

Protein-protein interactions (PPIs) are the basis of many biological processes. Therefore, a thorough understanding of PPIs could lead to a deeper comprehension of the cell physiology in both normal and disease states, facilitating relevant tasks like drug target developing and therapy design \cite{drugtarget, networkmedicine, drugrepurposing, drugrepositioning}. Consequently, PPIs prediction, that is the prediction of new protein-protein interactions given a Protein-protein Interaction Network, became crucial, both to investigate new associations among proteins and to check how these associations could contribute to the discovery of new methods in the biomedical field. 

High-throughput technologies\footnote{\url{https://en.wikipedia.org/wiki/Protein\%E2\%80\%93protein_interaction_screening}}, like Yeast two-Hybrid screens \cite{y2h}, and literature analysis have been used to generate a massive amount of data. Two of the most used high-throughput methods to determine protein-protein interactions are Yeast two-hybrid (Y2H) screens and Affinity-Purification Mass-Spectrometry (AP-MS). The main difference between the two is that Y2H is used to identify interactions between pair of proteins (binary interactions) while AP-MS \textit{identifies members of stable complexes, whether they directly interact with the bait protein or not}\footnote{\url{https://web.science.uu.nl/developmentalbiology/boxem/interaction_mapping.html}}. Despite the wide use and the effectiveness of the two aforementioned methods, they are {\bf very  expensive and time consuming}, revealing the need of high-throughput computational methods able generate high quality PPIs predictions. 

However, there are several problems that make this task difficult to solve, for example: 

\begin{enumerate}
    \item {\bf Absence of explicit encoding of negative knowledge}. This problem arises from the fact that generally only the positive interactions are reported in the medical literature and consequently contained in the protein-protein interactions databases. However, also negative interactions are equally useful when it comes to the training/testing of the PPIs prediction systems. Despite its importance, the lack of "gold standards" regarding the negative knowledge is often underestimated by the PPIs prediction systems, leading to biased performances and metrics. This issue will be discussed in more details in chapters \ref{chapter:negative_interactions_problem} and \ref{chapter:negative_methods_comparison}.
    
    \item {\bf Human PIN high incompleteness}. This is a problem associated with the small number of known protein-protein interactions in the human PIN \cite{estimated_size_interactome, Venkatesan}. One of the consequences of this problem is that all the methods that try to predict new PPIs are strongly limited by the lack of training data.
    
    In fact, according to \cite{booknetwork}: {\em Although computational efforts can mitigate the shortfall of literature-curated efforts and involve minimal experimental cost, any computational effort would be restricted by current limited knowledge of biological systems}.
\end{enumerate}

Despite the above mentioned problems, the number of systems for PPIs prediction is huge and constantly growing. These methods can be broadly divided in two categories: connection-based methods and protein-based methods.

\section{Connection-based methods}
\label{section:connection_based_methods}

These are the methods that leverage the structural information of the graph. 

\begin{figure}[]
    \centering
    \includegraphics[width=0.85\textwidth]{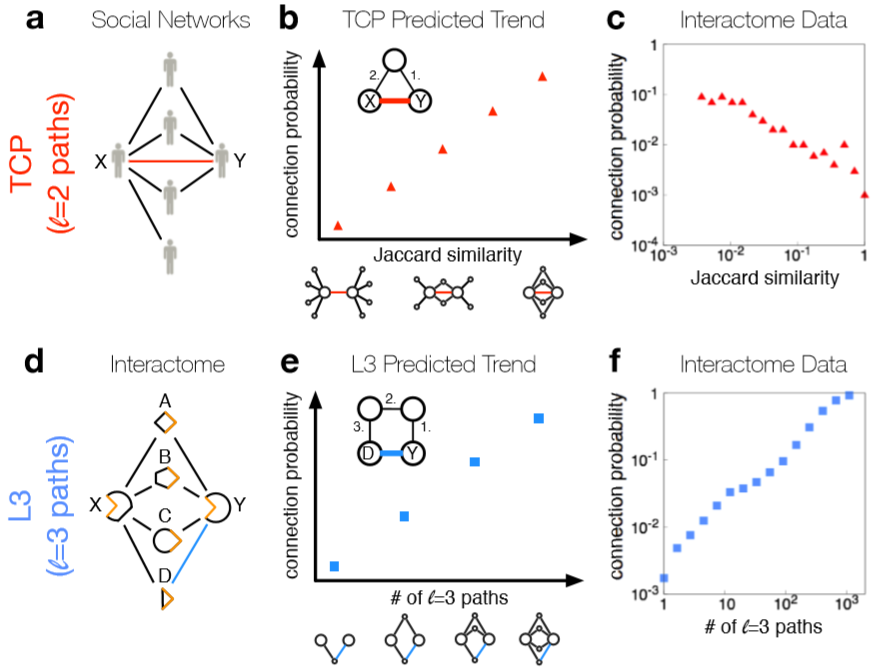}
    \caption{The comparison between the TCP and the L3 principle. Although the TCP is very effective in social networks (a), it is not as successful in the protein interaction networks, since the connection probability does not increase with the Jaccard similarity, as might be expected (b, c).
    Instead, the L3 principle is effective in the protein interaction networks, since the connection probability increases with the number of L3 paths (d, e, f) \cite{barabasippipred}.}
    \label{fig:l3vstcp}
\end{figure}

For example, the authors of \cite{barabasippipred} have shown that the Triadic Closure Principle (TCP, i.e. that two proteins likely interact if they share multiple interaction partners) is not valid for most protein pairs. Indeed, from a biological point of view, the interaction of two proteins \textit{y} and \textit{z} with an high number of proteins $x_1, x_2, .. , x_n$ means that \textit{y} and \textit{z} probably have higher interaction profile similarities. Therefore, since their interaction interfaces are similar, as opposed to complementary, they will tend not to interact \cite{barabasippipred, huri3}.  Instead of TCP, they proposed a link prediction principle based on paths of length three (L3 principle, figure \ref{fig:l3vstcp}). Specifically, they expect the interaction probability between two proteins A and B to be positively correlated with the number of paths of length three linking A to B, normalizing this number to avoid that hubs, since they introduce shortcuts in the network, could bias the result. 

Another example of a connection-based method for PPI prediction is the one presented in \cite{rwppipred}. The idea behind this method is that {\em two nodes having similar 'distances' to all other nodes in the network can potentially interact with each other}. To correctly compute the distances, they presented a novel random walk algorithm based on two ideas: (i) a small amount of resistance is added to each edge of the network to encourage the random walker to stay close to the starting point, and (ii) additional resistance is added to discourage the random walk from visiting a new node. The latter condition is introduced to avoid reaching, especially from hubs, nodes that may not be functionally related to the starting node. The result of this procedure is a $ |V| \times |V|$ probability matrix, where $|V|$ is the number of nodes. Finally, the Pearson correlation between two rows or columns is computed and the pairs of nodes having the correlation above the threshold are considered to be interacting.

However, the connection-based methods are strongly limited by data incompleteness, since, as explained above, only a small percentage of the total number of protein-protein interactions is known.  Furthermore, since these methods explore only network-related information, they are generally not able to properly predict interactions between proteins without known links. For example, applying the link prediction principle discussed in \cite{barabasippipred} (L3 principle), is not possible to predict an interaction between pairs of proteins not connected by any path of length three. Hence, the connections-based methods should be integrated with other methods using the proteins information (such as the amino acids sequence) to overcome the issue of predicting interactions between proteins not related by any topological property.

\section{Protein-based methods}
\label{section:protein_methods}

This category consists of all the methods that leverage the proteins information in order to predict new protein-protein interactions.

Specifically, the majority of these methods rely on the proteins' sequences information in order to predict a score for the probability of their interaction \cite{CNNRandomModule, cnnlstm, stackedAE, wavelet, negativeppi}. Indeed, the amino acids sequence of a specific protein can contain relevant information about the protein's structure and is therefore fundamental to identify possible interactions.

Generally, the protein-based methods first construct enriched representations of the proteins. For example, the authors of \cite{cnnlstm}, inspired by the "word2vec" model in Natural Language Processing, created an embedding for each amino acid, considering proteins' sequences as documents and amino acids as words. Then, they constructed an enriched representation of a protein by concatenating the embeddings of the amino acids in its sequence. 

\begin{figure}[]
    \centering
    \includegraphics[width=0.70\textwidth]{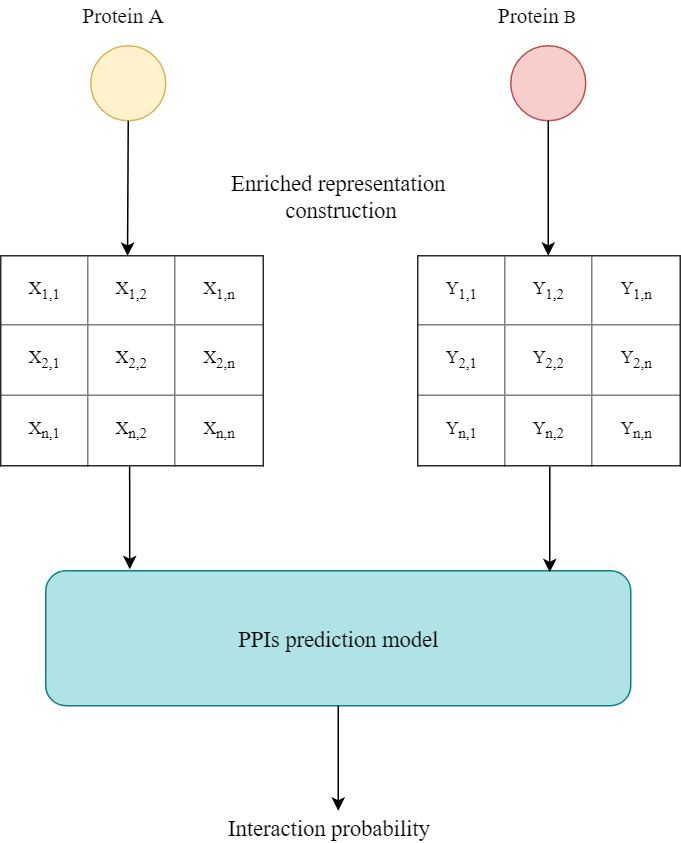}
    \caption{The typical flow of a PPIs prediction system. First, the two proteins of an interaction are represented in a more complex way, trying to capture their primary characteristics. Then, these representations are given as input to a PPIs prediction model (i.e., usually, a machine learning model) which outputs the probability of their interaction.}
    \label{fig:PPI_prediction_models}
\end{figure}

After their construction, these more detailed representations of the proteins are given as input to specific models (e.g. deep neural networks) to determine whether the two input proteins could interact or not. 

For example, the authors of \cite{stackedAE} used a stacked autoencoder, giving as input proteins coded into two different ways: Autocovariance method (AC) and Conjoint Triad method (CT). The AC method \cite{autocovariance} is a way to transform a sequence of amino acids into a $ N \times 7 $ matrix, where 7 is the number of physicochemical properties of amino acids that can reflect the various modes of a protein-protein interaction (hydrophilicity, volumes of sidechainsofaminoacids, polarity, polarizability, solvent-accessible surface area, hydrophobicity,  net charge index of side chains), and \textit{N} is the lag used by the formula, which is usually set as 30. Instead, in the Conjoint Triad method \cite{conjointtriad}, the 20 amino acids are firstly clustered into seven groups, based on their side chain volumes and their dipole. Then, after replacing each amino acid with the number of its cluster, a window composed of three amino acids is used to slide across the sequence, capturing the information about the frequency of each possible combination of three numbers (from 111 to 777).

Other examples of deep learning models used for PPIs prediction are those in \cite{cnnlstm} and \cite{CNNRandomModule}, that are, respectively, a CNN followed by a LSTM and a CNN followed by a random projection module, that {\em helps the model to investigate the combination of the patterns learned from two inputs proteins} and enables the model to ignore the order of the input profiles. An example of the functioning of a typical PPIs prediction system is given in Figure \ref{fig:PPI_prediction_models}.

Although the majority of these methods claim to achieve very high accuracies both on the training sets and on the validation sets, they should be tested under different datasets and conditions to better understand their predictive power. For example, the authors of \cite{cnnlstm} obtained excellent results (accuracy from 92\% to 98\%) on six different \textbf{positive} validation sets. However, since they used sets containing \textbf{only} positive examples, their performances could be incorrect since their model could be inclined to predict a new interaction as positive (a model predicting always "positive" would have a perfect accuracy). This hypothesis was also confirmed when we replicated their model (obtaining comparable results on the external datasets), and tested in on a dataset containing only negative examples \cite{negatome}, obtaining very poor results (about 30\% accuracy). One of the main reasons for this performance drop could be related to the bias induced by the choice of negative examples in the training phase.

\chapter{The negative interactions issue}
\label{chapter:negative_interactions_problem}

As mentioned in the previous chapters, the positive interactions are generally reported in the medical literature and therefore contained in all the databases of protein-protein interactions. Yet, it is known that the number of positive interactions in the human PIN is orders of magnitude lower than the number of negative interactions, since the PIN has a very low density (probably less than 0.005) \cite{estimated_size_interactome, Venkatesan}.

Still, the knowledge about negative protein-protein interactions (hereafter, NPIs) is {\bf very limited}, since they are generally considered less valuable than the positive ones. Despite this, high-quality NPIs are very important for several reasons:

\begin{enumerate}
    \item As previously mentioned, machine learning (and, recently, especially deep learning) has been widely used to build models designed to discover new PPIs. However, these models need to be trained and tested also on high-quality negative interactions, to prevent bias and to ensure that the performances and metrics reported are not distorted by the choice of inappropriate training and testing negative instances. 
    
    For example, analyzing several deep learning models, it seems like, since specific negative interactions are used for training, they only learn to distinguish the types of negative interactions in the training from the types of positive interactions in the training. Consequently, they will have very high accuracy on the training and test sets (since these datasets will only contain those types of negative and positive interactions), while they will obtain very poor results when tested on datasets containing positive and negative interactions of different types. This hypothesis is also confirmed in section \ref{section:experiments_ppi_models}, where some of these models will be analyzed in more detail. 
    \item The availability of negative knowledge, when very reliable, could help in deciding which interactions should not be tested in clinical experiments, speeding up the process of determining new protein-protein interactions. For example, an experimentally validated NPI between two proteins, could be useful to avoid testing this interaction again in the future.
    \item They could also be useful for the training of specific text mining models \cite{deepdive}. For example, they are required when {\em distant supervision}, which is a technique adopted to automatically label some training instances, based on some independent data sources, is used \cite{distantsupervision}.
\end{enumerate}

Despite their importance, currently there are no \textbf{reliable} datasets of negative interactions, causing a lack of negative "gold standards". Hence, for the training and testing of the models, NPIs are usually chosen using two different approaches: random sampling and different subcellular location.

\section{Random sampling} 
\label{section:random_sampling}

    In this approach, the negative instances are chosen by randomly pairing proteins and then removing the pairs already included in the positive examples. 
    
    Although this approach is currently considered as the most reliable for generating negative instances and it is widely used by the PPIs prediction systems, it presents several disadvantages.
    
    One disadvantage of this method is that, since the human PIN is highly incomplete and we do not know about the majority of the positive interactions, we could choose, as negative examples, interactions that could exist in reality, and this will add noise to the training set. However, it should be noted that, because of the low density of the human PIN, the probability that a negative interaction chosen by the random sampling approach is, in reality, a positive interaction, is very low (probably $<$ 0.5\%). 
    
    Another disadvantage is that the NPIs generated by this approach are {\bf highly influenced by the presence of hubs} in the set of positive PPIs. Indeed, since they are chosen randomly from the set of non-positive interactions, the probability that a NPI contains a hub belonging to the set of positive interactions is much lower than the probability that it contains a protein having a low degree in the positive set (since, for the way random sampling works, we will be more likely to choose an interaction between proteins having low centrality degrees). The influence of hubs is also reported by Yu et al. \cite{sequencebaseddonot}, who showed how the evaluation performances of different systems drop when it is used, as test set, a negative subset generated by \textit{balanced random sampling}, i.e. a negative set where each protein appears as many times as it does in the positive set. 
    
    However, later, Park et al. \cite{revisiting} demonstrated that the NPIs dataset generated using balanced random sampling is not suitable for testing purposes, since it differs from the population-level negative subset (i.e. the real negative subset) more than a negative dataset generated by simple random sampling. On the other hand, in line with Yu et al. \cite{sequencebaseddonot}, they confirmed how, when the negative instances generated from random sampling are used to train a model that uses the sequence of the protein as a feature, the bias induced by the presence of hubs in the positive set contribute, \textbf{in an inappropriate way}, to increase the performances. Therefore, balanced random sampling, since it maintains the degree of each proteins and therefore is not influenced by the presence of hubs, might still be appropriate for training purposes.
    
    Consequently, they suggested two different types of negative subset sampling:
    
    \begin{enumerate}
    \item The subset sampling used for cross-validated testing, where one prefers unbiased subsets so that the estimate of predictive performance can be safely assumed to generalize to the population level. For this type of task, the random sampling is more suitable than the balanced random sampling, since it is less biased and closer to reality. 
    
    \item The subset sampling used for training, where one desires the subsets that best train predictive algorithms, even if these subsets are biased. For this type of task, the balanced random sampling is more suitable than the simple random sampling, since its characteristics allow him to avoid the issue regarding the presence of hubs.
    
    \end{enumerate}

    Moreover, the negative protein-protein interactions generated by random sampling are not appropriate for specific biological contexts, where the interaction probability between proteins is higher than average, as shown in \cite{negy2h}. 

\section{Subcellular location heuristic}
\label{section:different_subcellular_location}

    In this approach, the negative instances are chosen by randomly pairing proteins in different subcellular locations. Hence, in the negative set generated by this approach there could not be interactions between two proteins in the Nucleus or two proteins in the Cytoplasm. 
    
    However, although the subcellular location constraint obviously reduces the number of false negatives, the authors of \cite{subcellular} have shown how this method is \textbf{highly affected by bias}. Specifically, they showed how the accuracy of a classifier depends on the co-localization threshold (i.e. the allowed similarity between the cellular compartments) of pairs of proteins in the negative examples. Indeed, the accuracy is higher for low co-localization threshold (i.e. using, as negative examples, pairs of proteins with a strong difference in their subcellular localization), and this can be explained {\em by the fact that the constraint on localization restricts the negative examples to a subspace of sequence space, making the learning problem easier than when there is no constraint}. 
    
    Furthermore, confirming the previous statement, Zhang et al. \cite{negativeppi} trained the same model using, as negative examples, first, pairs of proteins in different subcellular locations and next, pairs of proteins retrieved with different methods (like sequence similarity and length of the shortest paths, as will be explained later). As a result, the former model (i.e. the one trained using negative interactions derived from different subcellular locations) performed better on datasets composed  only of positive interactions (96\% vs 86\%) but much worse on datasets composed  only of negative interactions (4\% vs 17\%), showing that, when the negative interactions generated from the different subcellular location strategy are used for training, the model is inclined to predict a new protein pair as a positive interaction. 
    
    Albeit it has been shown that this approach is highly biased, it is still used in several recent works \cite{cnnlstm, stackedAE, wavelet}.

\section{Other methods}
\label{section:other_negative_methods}
Motivated by the limitations of the existing solutions, a research area has been focused on how to produce reliable negative examples. The following subsections contain an analysis regarding the methods proposed.

\subsection{Bait-prey approach}
\label{subsection:bait_prey_approach}

\begin{figure}[]
    \centering
    \includegraphics[width=0.90\textwidth]{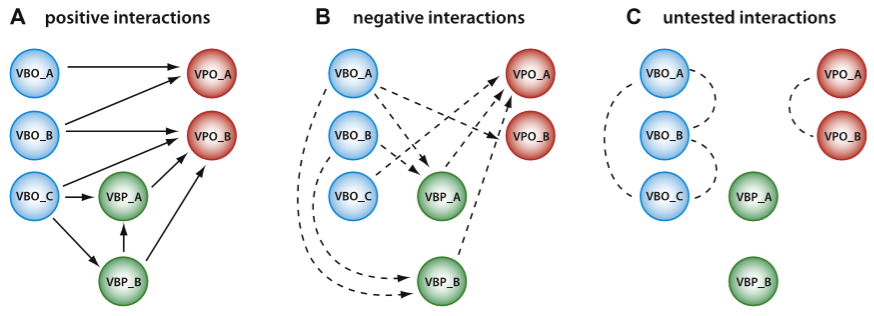}
    \caption{The functioning of the viability analysis \cite{negy2h}. 
    (A) the positive interactions retrieved by the large-scale yeast two-hybrid screens. (B) The negative interactions generated are those from proteins acting as baits to proteins acting as preys. (C) each link between two proteins acting only as baits (two VBO) or between two proteins acting only as preys is not considered as negative interaction, since it is considered to be untested.}
    \label{fig:neg_y2h}
\end{figure}

Trabuco et al. \cite{negy2h} proposed a method that generates reliable negative examples using viability analysis\footnote{\url{http://www.russelllab.org/negatives/}}. To do this, they retrieved the proteins in the interactions of the IntAct dataset that were derived from large-scale Y2H experiments, and divided them into 2 categories: viable baits (VB, i.e. proteins acting as baits for at least one interaction) and viable prey (VP, similarly defined). Then, a protein is considered a viable bait only (VBO) if it is a VB but not a VP.
Viable prey only (VPO) are similarly defined. Each link from a VB to a VP that is not present in the dataset of positive interactions is considered as a negative interaction, since {\em it is assumed to have been tested} (for the way the large-scale Yeast two-hybrid works). Instead, each link between two VBO or between two VPO is {\em deemed untested}, since the proteins should not have been tested in any way. This method is presented in Figure \ref{fig:neg_y2h}.

In order to generate more reliable datasets, the authors also proposed a variant of their model using a shortest path heuristic, based on the fact that \textit{due to the small-world property exhibited by biological networks \cite{networkmedicine}, interacting proteins are expected to be near each other in the network}. In this variant they removed from the negative set generated a negative interaction between two proteins A and B if the shortest path from A to B in the PIN was less than a certain threshold.
\\
Altough the bait-prey method has shown improvements over the random sampling on the protein-protein interactions networks of different species and organisms (e.g. yeast), especially when the shortest path variant was used, it did not show clear benefits when applied to the human PIN, where the negative interactions generated using this approach were similar (in terms of performances) to those generated by random sampling. 

This could be partly caused by the human PIN incompleteness issue.

\subsection{Negatome 2.0}
\label{subsection:negatome}
One of the few negative datasets that have been generated is the Negatome 2.0 \cite{negatome}, which is a database of proteins and protein domains that are unlikely to engage in physical interactions\footnote{\url{http://mips.helmholtz-muenchen.de/proj/ppi/negatome/}}. 

The negative interactions contained in this database are retrieved using 2 methods: (i) analysing, through text mining techniques, a corpus of medical articles, and (ii) analysing the three-dimensional structures of the proteins. 

A first step that is mandatory when trying to retrieve candidate interactions from the medical literature is that of finding all the sentences that might contain an interaction, to remarkably decrease the execution time of the method. To address this problem, the developers used a specific tool, Excerbt \cite{excerbt_senna}, to retrieve only the sentences where: (i) the proteins are both the agent (the entity that carries out the action of the verb) and the theme (the entity that receives the action of the verb), (ii) there is a verb referring to interactions or binding and (iii) there is a negation. 

Furthermore, the authors defined a confidence score based on simple features, like the length of the sentence and the type of the relation (some relations are considered to be stronger than others), to assess the precision of the linguistic analysis. Then, the precision and the recall of the resulted "non-interactions" have been evaluated through manual validation. 

Regarding the precision, the authors analyzed a sample of the non-interactions generated, finding that more than 50\% of them classified correctly. In addition, it also turns out that the confidence score was very informative about the annotation quality. Indeed, {\em among the 20 top scoring sentences the precision of text mining was 95\%, while for the median 20 and the bottom 20 sentences it was 45\% and 15\%, respectively}. Despite the importance of the confidence score, it is not released by the authors of the dataset, and it is considered to be internal to the procedure. 

To compute the recall, i.e. how many of the non-interactions described in the literature were found by Excerbt\footnote{\url{http://mips.helmholtz-muenchen.de/excerbt/}}, they investigated how well Negatome 1.0 (a small database derived by the manual curation of the literature) could be reproduced by text mining. Regarding this measure, the results are unsatisfactory. Indeed, when they analyzed a sample of 20 non-interactions from Negatome 1.0, they found out that only five non-interactions were "reachable" from Excerbt, and only 3 of them were correctly identified. Among the others 15 non-interactions, one was misclassified by Negatome 1.0, and the remaining 14 could not be found by Excerbt, since they were present in sentences containing particolar grammar features (e.g. ellipsis, anaphora) or protein names not covered by the Excerbt ontology. 

Overall, the system exhibits a very low recall and a not good enough precision. Moreover, considering the number of positive interactions already known and the fact that negative interactions should be considerably more than the positives, the number of non-interactions contained in Negatome is very small, less than 2000. 

\subsection{Sequence similarity and shortest paths}
\label{subsection:zhang_negative}
Very recently, Zhang et al. presented two novel methods for generating negative protein-protein interactions \cite{negativeppi}. 

The first one selects pairs of proteins having a lower sequence similarity score, based on the idea that \textit{for an experimentally validated PPI between protein i and j, if a protein k is dissimilar to i, there is a low possibility that k interacts with j}. To do this, they first calculated the sequence similarity score using the BLOSUM50 matrix, and then they normalized the score using the following formula:\\

\begin{equation}
    \widetilde{bl}\left ( i, j \right ) = \frac{bl\left ( i, j \right ) - min\left \{ bl\left ( i, 1 \right ), ... , bl\left ( i, n \right ) \right \}}{max\left \{ bl\left ( i, 1 \right ), ... , bl\left ( i, n \right ) \right \}}
\end{equation}
\\

Where n represents the number of proteins and $bl\left ( i, j \right )$ is the score of the BLOSUM50 matrix for proteins $i$ and $j$.

However, Luck et al. \cite{huri3} have shown that, even if global sequence identity is indicative of shared interaction interfaces, {\em it likely fails to identify pairs of proteins whose shared interaction interface is small} and that {\em the functional relationships between proteins are not necessarily identified by sequence identity}. 

Instead, the second method selects negative interactions based on the following observation: {\em the probability, for a pair of proteins, to share similar functions (so to interact) reduces with the increase of the length of the (shortest) path between the two proteins}. However, as also mentioned in section 3.1, the authors of \cite{barabasippipred} have shown that the principle that two proteins are likely to interact if they share multiple interaction partners, that can also be called L2, is not valid for PPI networks. Furthermore, they have shown that the connection probability varies using different path lengths, demonstrating that there is no  clear correlation. Hence, the assumption on which the second method of negative interactions generation is based (i.e. that two proteins are more likely to interact if the path between them is shorter), could be partly unfounded.

\subsection{Hybrid methods}
Finally, there are some methods that use cellular compartment information (CCI) and other information to generate high-quality NPIs, stating that the two proteins in each pair should not have overlap in any of these areas (i.e. they should have different cellular compartment, different functions, etc.) \cite{homologousneg, detectingnip}. Yet, if only the use of different CCI led to biased results since the constraint makes the task easier and the performance biased, as shown in \cite{subcellular}, imposing a filter using the intersection between CCI and other constraints, even though it obviously reduces false positives even more, could worsen the problem.

\chapter{Comparison of methods generating negative interactions}
\label{chapter:negative_methods_comparison}

We performed several experiments in order to assess the reliability of the negative interactions generated by two of the previously explained methods: random sampling (section \ref{section:random_sampling}) and bait-prey (subsection \ref{subsection:bait_prey_approach}, \cite{negy2h}). The former was chosen because it is widely used and surely less biased than the different subcellular location approach, and the latter was chosen because, although not extensively used, its premises classify it as one of the most reliable methods among those described in chapter \ref{chapter:negative_interactions_problem}. 

Both of these methods, to generate negative interactions, need a training set of positive interactions, from which the set of proteins will be extracted. For this purpose, we decided to consider the 5393 interactions contained both in HI-14 and HI-19 as high-quality positive interactions, and, from now on, they will be referred as HI-19-TRAIN. 

First, we considered the set $P$ of proteins in HI-19-TRAIN (2642 different proteins)\footnote{It should be noted that the proteins in HI-19-TRAIN are only a subset of the proteins in the intersection between HI-14 and HI-19. Indeed, a protein present in both datasets that is not contained in a common interaction is not present in HI-19-TRAIN.}.  

Using the set of 5393 positive interactions in HI-19-TRAIN (i.e. the interactions between pairs in $P$ which are in HI-14 and are also present in HI-19), different sets of negative interactions, $GNI(P)$, were generated, using the random sampling approach, the bait-prey, and some variations of the latter. 

One of the available ways to test the quality of the negative interactions generated by these methods is to calculate the number false negative interactions (i.e. predicted negative interactions that were instead positive).

Since these methods, for how they work, can only generate negative interactions between proteins in P, the positive interactions that can be used as "ground truth" are only those \textbf{between the same set of proteins P}, that are in HI-19 (that also represents our test set) and are not in HI-14 (since, if they are both in HI-19 and in HI-14, they would be contained in the training set HI-19-TRAIN). These positive interactions, that represent our test set, will be, from now on, referred as HI-19-TEST, and their number is 10358 (|HI-19-TEST|$=10358$). 

For clarification, figure \ref{fig:neg_pos_ppi} presents a graphical view of all the possible interactions between a set of proteins P.

In this figure:

\begin{enumerate}
    \item $NI(P)$ are all the existing  negative interactions among pairs of proteins in P. 
    \item $PI(P)$ are all the existing  positive interactions among pairs of proteins in P.
    \item $KPI(P)$ are the known (tested) positive interactions between proteins in P. In our case, $KPI(P)$ is represented by the interactions in HI-19 between proteins in P. Each of these interactions will be in HI-19-TRAIN if it is contained also in HI-14 and will be in HI-19-TEST otherwise.
    \item $KNI(P)$ are the known (tested) negative interactions between the proteins in P. In our case, $KNI(P)$ is represented by the interactions in HQND between proteins in P.
    \item $GNI(P)$ are the generated negative interactions between proteins in $P$. The negative interactions generated by a specific method $m$ (see Chapter \ref{chapter:negative_interactions_problem}) will be referred as $GNI_m(P)$.
\end{enumerate}

On the basis of these sets, several evaluation metrics were defined: Verified Error Rate, Minimum bound of Error Rate, Verified Success Rate and Approximated True Error Rate.




\section{VER and MER}
\label{section:VER_and_MER}

\begin{figure}[]
    \centering
    \includegraphics[width=0.90\textwidth]{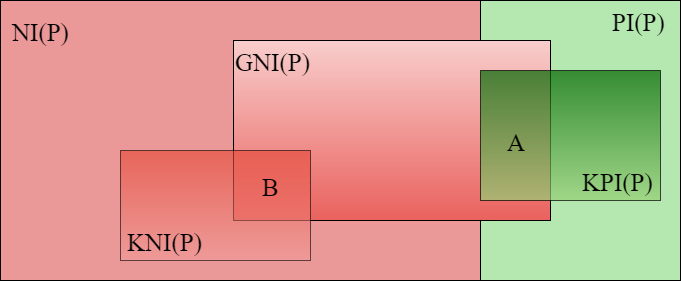}
    \caption{Diagram of positive and negative interactions between a set of proteins $P$. $NI(P)$ are the negative interactions between proteins in P, whereas $PI(P)$ are the positive ones. $KNI(P)$ are the known negative interactions between proteins in P, which, in our case, are a subset of all the negative interactions in HQND. Instead, $KPI(P)$ are the known positive interactions between proteins in P, that are, in our case, a subset of the interactions contained in HI-19. Each method $m$ generates a set of negative interactions, in the figure denoted as $GNI(P)$. Finally, A and B are, respectively, the negative interaction generated which are known to be positive and those which are known to be negative. Please note that the proportions of the sets in the figure do not reflect reality. }
    \label{fig:neg_pos_ppi}
\end{figure}

Considering the sets in figure \ref{fig:neg_pos_ppi}, the Verified Error Rate (VER) of a specific method $m$ can be defined as: \\

\begin{equation}
    VER=\frac{GNI_m\left ( P \right ) \cap KPI\left ( P \right )}{KPI\left ( P \right )}
\end{equation}
\\

That is the fraction of known positive interactions ($KPI(P)$) that were generated as negative interactions by the method $m$ ($GNI_m(P)$). This measure quantifies how good the method is at avoiding the generation of positive interactions with respect to the number of positive interactions known.

Obviously, this measure is strictly related to the number of interactions generated by the method, since, for example, a simple method not generating any negative instance would have a perfect score (0\% VER). 

Instead, the Min. bound of Error Rate (MER) of a specific method $m$ can be defined as: \\
    
\begin{equation}
    MER=\frac{GNI_m\left ( P \right ) \cap KPI\left ( P \right )}{GNI\left ( P \right )}
\end{equation}
\\   

That is the fraction of the negative interactions generated that were instead positive (false negatives). This measure defines a lower bound on the real error rate of the negative interactions generated. 

%
%
    




We generated negative interactions with four different methods, and the overall results are presented in Table \ref{table:VER_MER_HI19}. 

As first method, the \textbf{bait-prey approach} \cite{negy2h} was used to obtain about 2.32 million negative interactions. The interactions that turned out to be incorrect (based on the "ground truth" of the 10358 interactions in HI-19-TEST) were \textbf{8618} (83\% VER). In relation to the total number of negative interactions generated, the wrong ones represent the \textbf{0.37\%} (MER). 

In the same way, we generated 2.32 million negative interactions using the \textbf{random sampling approach}\footnote{Note that, differently from the bait-prey approach, the number of negative interactions generated by the random sampling approach must be set.}. The interactions that turned out to be incorrect were \textbf{6856} (66.2\% VER). In relation to the total number of generated negative interactions, we also obtained better results with respect to the bait-prey approach, since the wrong interactions represent the \textbf{0.294\%} (MER) of the total number of negative interactions generated. For the negative interactions generated by the random sampling approach, since they are randomly chosen, the experiment has been repeated 50 times, taking the average of the wrong interactions at the end. Given how random sampling works, the number of generated interactions does not affect the MER (i.e. it can generate a different number of interactions maintaining a similar MER). However, the VER, as described previously, is highly dependent on the number of interactions generated. Therefore, the experiment with the random sampling approach was repeated several times with different sizes, considering the number of interactions generated by the three other methods, as reported in table \ref{table:VER_MER_HI19}.

\begin{table}[]
\centering
\begin{tabular}{|c|c|c|c|c|}
\hline
Method & \begin{tabular}[c]{@{}c@{}}Generated\\ NI\end{tabular} & \begin{tabular}[c]{@{}c@{}}Wrong\\ Interactions\end{tabular} & \begin{tabular}[c]{@{}c@{}}Verified Error\\ Rate (VER)\end{tabular} & \begin{tabular}[c]{@{}c@{}}Minimum Error\\ Rate (MER)\end{tabular} \\ \hline
Bait-prey & 2.32 million & 8618 & 83\% & 0.37\% \\ \hline
\multirow{3}{*}{Random sampling} & 2.32 million & 6856 $\pm$ 55 & 66.2\% $\pm$ 0.5 & 0.294\% $\pm$ 0.002 \\ \cline{2-5} 
 & 785 thousand & 2318 $\pm$ 17 & 22.38\% $\pm$ 0.16 & 0.295\% $\pm$ 0.002  \\ \cline{2-5}
 & 705 thousand & 2077 $\pm$ 15 & 20.05\% $\pm$ 0.14 & 0.294\% $\pm$ 0.002 \\ \hline
\begin{tabular}[c]{@{}c@{}}Modified\\ Bait-prey\end{tabular} & 785 thousand & 2625 & 25.3\% & 0.33\% \\ \hline
\begin{tabular}[c]{@{}c@{}}Modified \&\\ Filtered\\ Bait-prey\end{tabular} & 705 thousand & \textbf{1067} & \textbf{10.3\%} & \textbf{0.15\%} \\ \hline
\end{tabular}
\caption{Comparison of the error rates on the HI-19 test set. The best results are obtained by the MFB-P method, which generates a set of interactions in which only the 0.15\% are wrong.}
\label{table:VER_MER_HI19}
\end{table}

Then, trying to improve the bait-prey approach, we considered only the interactions from Viable Bait Only to Viable Prey Only (i.e. a subset of the interactions generated by the original bait-prey method). This modification, from now on referred as Modified Bait-Prey (MB-P) method, generated 785 thousand negative interactions, of which \textbf{2625} were wrong (25.3\% VER). In relation to the total number of negative interactions generated, the wrong ones represent the \textbf{0.33\%} (MER). Hence, the change proposed led to an increase in the quality of the interactions generated with respect to the original bait-prey approach. Nonetheless, MER and VER are still higher than those of the random sampling approach (0.33\% vs 0.294\% and 25.3\% vs 22.38\%) and this difference turned out, after a \textit{one-tail test}, to be \textit{statistically significant} (p < .00001).

Finally, as also suggested in the bait-prey paper as a possible extension of the proposed method, we decided to filter the interactions generated by the previous method (MB-P) using the shortest path heuristic. Specifically, each negative interaction (A, B) generated is removed if the length of the shortest path between A and B in the Protein Interaction Network constructed from the interactions in HI-19-TRAIN is less or equal than three, based on the results in   \cite{barabasippipred}. This method, from now on, will be referred as Modified and \textbf{Filtered} Bait-Prey (MFB-P) method.

From a set of over 785 thousand interactions, 705 thousand interactions remained. Among these interactions, only \textbf{1067} were wrong (10.3\% VER). Similarly, the MER decreased to \textbf{0.15\%}, achieving for both measures a better quality with respect to all the other methods. 

\begin{table}[]
\centering
\begin{tabular}{|c|c|c|c|c|}
\hline
Dataset & \begin{tabular}[c]{@{}c@{}}Number of\\ interactions\end{tabular} & \begin{tabular}[c]{@{}c@{}}Number of\\ proteins\end{tabular} & \begin{tabular}[c]{@{}c@{}}Interactions\\ in KPI(P)\end{tabular} & \begin{tabular}[c]{@{}c@{}}Density \\ of KPI(P)\end{tabular} \\ \hline
HI-19 & 54496 & 8490 & 10358 & 0.004 \\ \hline
LIT-BM & 13030 & 6047 & 131 & 0.011 \\ \hline
\end{tabular}
\caption{Comparison between the HI-19 and the LIT-BM datasets. With respect to HI-19 test set, the LIT-BM test set has a considerably smaller number of interactions but a higher density.}
\label{table:comparison_LIT-BM_hi19}
\end{table}

To further define the VER of the MFB-P method, we also carried out an analysis to identify the VER confidence intervals. This analysis allows us to compute an interval (composed of a lower bound \textit{LB} and an upper bound \textit{UB}) in which the examined value will be with a specific probability N, when the model will be tested on other datasets. As a result, with \textbf{95\%} (N) probability, the VER of the last approach will be between \textbf{9.7\%} (\textit{LB}) and \textbf{10.9\%} (\textit{UB}) also when different test sets will be used. 

Moreover, to further assess the reliability of the negative interactions generated and to compare the approaches in a different setting, the same four methods were also tested on LIT-BM, a literature curated dataset (see Section \ref{section:lit-bm}). 

As said in chapter \ref{chapter:datasets}, LIT-BM, although smaller, provides more information in the dense zone than HI-19, since it is biased towards most studied proteins, while HI-19 is more systematic and unbiased. As can be seen in Table \ref{table:comparison_LIT-BM_hi19}, which contains a comparison between the datasets, the density of the PIN constructed from the KPI(P) of LIT-BM (i.e. from the interactions between the set of proteins $P$ that are in LIT-BM and not in HI-19-TRAIN) is almost \textbf{three times higher} than the one of the PIN derived from the KPI(P) of HI-19 (i.e. HI-19-TEST). \\
The results of the experiment are shown in Table \ref{table:VER_MER_LIT-BM}.

\begin{table}[]
\centering
\begin{tabular}{|c|c|c|c|c|}
\hline
Method & \begin{tabular}[c]{@{}c@{}}Generated\\ NI\end{tabular} & \begin{tabular}[c]{@{}c@{}}Wrong\\ Interactions\end{tabular} & \begin{tabular}[c]{@{}c@{}}Verified Error\\ Rate (VER)\end{tabular} & \begin{tabular}[c]{@{}c@{}}Minimum Error\\ Rate (MER)\end{tabular} \\ \hline
Bait-prey & 2.32 million & 65 & 49.6\% & 0.0028\% \\ \hline
\multirow{3}{*}{Random sampling} & 2.32 million & 61 $\pm$ 5.6 & 46.9\% $\pm$ 4.2 & 0.0026\% $\pm$ 0.0002 \\ \cline{2-5} 
 & 785 thousand & 23 $\pm$ 2 & 17.5\% $\pm$ 1.5 & 0.0029\% $\pm$ 0.0002  \\ \cline{2-5} 
 & 705 thousand & 20 $\pm$ 2 & 15.2\% $\pm$ 1.5 & 0.0028\% $\pm$ 0.0002 \\ \hline
\begin{tabular}[c]{@{}c@{}}Modified\\ Bait-prey\end{tabular} & 785 thousand & 16 & 12.21\% & \textbf{0.002\%} \\ \hline
\begin{tabular}[c]{@{}c@{}}Modified \&\\ Filtered\\ Bait-prey\end{tabular} & 705 thousand & \textbf{14} & \textbf{10.68\%} & \textbf{0.00198\%} \\ \hline
\end{tabular}
\caption{Comparison of the error rates on the LIT-BM test set. Differently from the experiment on the HI-19 test set, the negative interactions generated by the random sampling approach perform even worse than the MB-P, showing their inadequacy when the density of the PIN is higher than average \cite{negy2h}.}
\label{table:VER_MER_LIT-BM}
\end{table}

As can be seen from the table, the situation is slightly different from the experiment on the HI-19 dataset (described in Table \ref{table:VER_MER_HI19}). Indeed, the results of this experiment show that, in this case, the performances of the random sampling method are also worse than those of the MB-P method. This fact is probably correlated with the higher density of the KPI(P) constructed from LIT-BM, and with the hypothesis that the random sampling approach performs bad when the interaction probability is higher than average, as explained in section \ref{section:random_sampling}. The negative interactions that perform best are still those generated by the MFB-P method, although the difference with the MB-P method, concerning the MER, is not \textit{statistically significant}. Moreover, the Verified Error Rate of the MFB-P method meets the requirements of the VER confidence intervals analysis previously done\footnote{Since it is between 9.7\% and 10.9\%.}, confirming the validity of the analysis. 

\section{VSR}
\label{section:VSR}
We used the High Quality Negatome Dataset (HQND, section \ref{section:HQND}) to check how many reliable negative interactions in this dataset were recovered from the four methods. This measure, called Verified Success Rate (VSR), can be defined, based on figure \ref{fig:neg_pos_ppi}, as:\\

\begin{equation}
    VSR=\frac{GNI_m\left ( P \right ) \cap KNI\left ( P \right )}{KNI\left ( P \right )}
\end{equation}
\\

Since, as explained previously, the four methods can only generate negative interactions with both proteins in the set P, we found that the maximum possible overlap between the negative interactions in HQND and the negative interactions generated by a specific method is 31, which corresponds to the number of interactions in HQND having both proteins in the set P ($|KNI(P)| = 31$). 

The results of the experiment are presented in table \ref{table:verified_success_rate}. As shown in the table, only 6 interactions were recovered from the bait-prey method, while the MB-P method and the MFB-P method both recovered 2 interactions. Instead, the random sampling method recovered, on average, 13 interactions when generating 2.32 million interactions, and 5 and 4 when generating, respectively, 785 thousand interactions and 705 thousand interactions. 

However, since the maximum possible number of interactions that could be recovered is very small (31), these results are not \textit{statistically significant}.

\section{ATER}
\label{section:ATER}

Although several models and their performances are compared using reliable training and test sets, the results of the previous experiments do not provide insights about the real confidence that we have in the negative interactions generated by the four methods. In other terms, we are currently unable to answer the following research question:  given an interaction generated by a specific method $m$, what is the probability that the  interaction is actually negative?

\begin{table}[]
\centering
\begin{tabular}{|c|c|c|c|}
\hline
Method & \begin{tabular}[c]{@{}c@{}}Generated\\ NI\end{tabular} & \begin{tabular}[c]{@{}c@{}}Recovered\\ NI\end{tabular} & \begin{tabular}[c]{@{}c@{}}\% of recovered\\ NI\end{tabular} \\ \hline
Bait-prey & 2.32 million & 6 & \textbf{19.3\%} \\ \hline
\multirow{3}{*}{Random sampling} & 2.32 million & 13 & \textbf{41.9\%} \\ \cline{2-4} 
 & 785 thousand & 5 & \textbf{16\%} \\ \cline{2-4} 
 & 705 thousand & 4 & \textbf{12\%} \\ \hline
\begin{tabular}[c]{@{}c@{}}Modified\\ Bait-prey\end{tabular} & 785 thousand & 2 & \textbf{6\%} \\ \hline
\begin{tabular}[c]{@{}c@{}}Modified \&\\ Filtered\\ Bait-prey\end{tabular} & 705 thousand & 2 & \textbf{6\%} \\ \hline
\end{tabular}
\caption{Negative interactions in HQND recovered from the four methods. In this case, the best performances are obtained by the random sampling approach, which, on average, recovered twice the number of interactions recovered from the other methods.}
\label{table:verified_success_rate}
\end{table}

As far as the previous experiments are concerned,  they provide a reliable estimate of the Verified Error rate (VER) and the Min. bound of Error Rate (MER), which measure the goodness of the entire set of negative interactions generated by a method. For example, we know that the original bait-prey method has a MER of 0.37\%, meaning that the 0.37\% of the negative interactions generated are instead positive. Consequently, given a specific interaction \textit{i} in the set of the negative interactions generated by the bait-prey method, the probability that it is wrong (i.e. $i$ is, in reality, a positive interaction) is 0.37\%, while the probability that it is right (i.e. $i$ is actually a negative interaction) is 99.67\%. 

However, this represents a \textbf{minimum bound} estimate of the error rate, because we do not know all the positive interactions between the set of 2642 proteins in the training set (in figure \ref{fig:neg_pos_ppi}, these interactions are represented as $PI(P)$). 

Nevertheless, we know that the density of the human PIN is \textasciitilde 0.003, considering 20'000 genes and 650'000 total interactions \cite{estimated_size_interactome}. Based on that, we can try to reconstruct the total number of interactions that there should exist between the 2642 proteins, i.e. the approximated cardinality of $PI(P)$. However, it should be noted that these 2642 proteins contain some of the most important and studied proteins, therefore the density in this PIN is probably higher than 0.003. Consequently, we considered it as 0.006, i.e. twice the density of the whole human PIN.

Mathematically speaking, we are looking for the unknown variable X in the following equation:\\

\begin{equation}
    \frac{2X}{2642 \times 2641} = 0.006
\end{equation}
\\

This corresponds to the number of interactions that there should be between the 2642 proteins to ensure that the density of this sub-PIN is 0.006.

The unknown X corresponds to the number \textbf{20932}, that, with respect to the number of interactions already known (10358) is \textbf{2.02} times bigger.

Hence, since we know that the bait-prey method has a MER of 0.37\% if we consider 10358 interactions, we can assume that it would have a MER of 0.747\% when considering all the 20392 interactions. This implies that each negative interaction generated by the bait-prey method is wrong with probability 0.747\% and right with probability 99.253\%.

It should be noted that now, more than a min. bound of the Error Rate, this measure represents an Approximated \textbf{True} Error Rate (ATER). 

In the same way, the ATER for each of the three remaining methods can be computed. The overall results are presented in table \ref{table:ATERs}.

\begin{table}[]
\centering
\begin{tabular}{|c|c|c|}
\hline
Method & \begin{tabular}[c]{@{}c@{}}Generated\\ NI\end{tabular} & \begin{tabular}[c]{@{}c@{}}Approximate True\\  Error Rate (ATER)\end{tabular} \\ \hline
Bait-prey & 2.32 million & 0.747\% \\ \hline
Random sampling & 2.32 million & 0.594\% \\ \hline
\begin{tabular}[c]{@{}c@{}}Modified\\ Bait-prey\end{tabular} & 785 thousand & 0.66\% \\ \hline
\begin{tabular}[c]{@{}c@{}}Modified \&\\ Filtered\\ Bait-prey\end{tabular} & 705 thousand & 0.3\% \\ \hline
\end{tabular}
\caption{ATERs of the various methods. These measures represent an approximation of the percentage of the wrong interactions generated.}
\label{table:ATERs}
\end{table}
\section{Final considerations}

Based on the experiments done, the following conclusions can be drawn:

\begin{itemize}
    \item The method that performs best is modified and filtered bait-prey (MFB-P) method, although the shortest path heuristic on which it is based could make the negative interactions dataset biased.
    \item The second best method, based on the experiment on the HI-19 test set (table \ref{table:VER_MER_HI19}), is random sampling. However, this method performed bad on the LIT-BM test set, confirming the hypothesis that, when trying to generate negative interactions from a set of positive interactions having a density higher than average, other methods must be used. 
    \item The ATER of the various methods is still low. As a result, whatever the method, the probability for a specific negative interaction generated to be \textbf{actually negative} is very high (from 99.25\% to 99.7\%). Therefore, the main issue to consider when generating negative interactions is to make sure that the generated dataset is not biased.
\end{itemize}

\chapter{Computational PPIs prediction}
\label{chapter:ppi_prediction_model}

As previously explained, the cost and time restrictions of high-throughput technologies used to experimentally discover new protein-protein interactions underlined the necessity to create new computational methods able to generate high quality PPIs predictions. 

In this chapter:

\begin{itemize}
    \item A machine learning model for protein-protein interactions prediction is presented. Therefore, the aim is to implement a system able to learn whether the two input proteins interact or not. 
    \item We compared, again, two of the methods for generating negative protein-protein interactions: bait-prey and random sampling. However, this time the \textbf{balanced} random sampling, i.e. a random sampling that generates a set of negative interactions in which each protein appears as many times as it does in the positive set, is used, since it is more appropriate for training \cite{revisiting}.
    Furthermore, this time, the negative interactions generated by these methods were compared based on how they affect the models' performances in the PPIs prediction task. 
    \item A set of features relevant to the protein-protein interactions prediction task are described and analyzed, also comparing their importance.
    \item The performances of several machine learning models are compared, both among themselves and with other two sequence-based state-of-the-art methods.
    \item We showed how two sequence-based state-of-the-art methods perform when more reliable experimental settings are considered.
    
\end{itemize}

The models will be trained and tested using four datasets. As training datasets, we used HI-19 (section \ref{section:huri_dataset}, \textbf{positive} dataset) and, one at the time, each of the datasets generated by the approaches compared in chapter \ref{chapter:negative_methods_comparison} (as \textbf{negative} datasets), that are:
\begin{enumerate}
    \item {\textbf{Original Bait-Prey (OBP) Dataset:}}\\
    This dataset is generated by the original bait-prey method. Since it has the highest Verified Error Rate (VER, 83\%) and the highest Minimum Error Rate (MER, 0.37\%), it is considered as the less reliable of the four. 
    \item {\textbf{Modified Bait-Prey (MB-P) Dataset:}}\\
    This dataset is generated from a variation of the original method, in which the negative interactions retrieved are only those from VBO (Viable Bait Only) to VPO (Viable Prey Only). Although it still has a relevant MER (0.33\%), it can be considered more reliable than the OBP dataset. 
    \item {\textbf{Modified and Filtered Bait-Prey (MFB-P) Dataset:}}\\
    This dataset is generated from a variation of the MB-P method, in which each negative interaction between two proteins A and B is removed if the shortest path between A and B in the Protein Interaction Network is less or equal than three. This dataset, having the minimum MER and VER, can be considered as the most reliable. However, since the shortest path heuristic removed all the negative interactions between relatively near (in the Protein Interaction Network) proteins, this dataset could be biased.
    \item {\textbf{Balanced random sampling Dataset:}}\\
    As described in section \ref{section:random_sampling}, one of the problems of the random sampling approach is that the presence of hubs (i.e. proteins with a lot of interactions) makes the training set biased and contributes, in an inappropriate way, to increase performances. Instead, in a negative dataset generated by \textbf{balanced} random sampling, the degree (i.e. the number of interactions) of each protein is maintained, so that each protein appears as many times as it does in the positive set.
\end{enumerate}

Regardless of the method adopted, the number of negative interactions in the training set was set to be equal to the number of positive interactions in the training set.

Instead, as validation sets, the HQND (\textbf{negative} dataset, section \ref{section:HQND}) and the LIT-BM dataset (\textbf{positive} dataset, section \ref{section:lit-bm}) were used. Although smaller than the training sets, these two highly reliable validation sets are of crucial importance to evaluate the models in the prediction of positive and negative interactions.
    
\section{Features}
\label{section:features}

    Feature selection/engineering is one of the most important steps in the creation of a machine learning model. Its aim is to identify the set of possible features that could contribute, in any way, to increasing the model's performance for a specific task (in this case, PPIs prediction). Then, all these features must be preprocessed in order to obtain the best possible configuration for the model. Indeed, the way in which we will choose to preprocess each feature will affect the model, based on the importance of the preprocessed feature with respect to the task that we would like to perform. 

Below is the list of all the features that were identified for the PPIs prediction task and how they were preprocessed:
    
    \begin{enumerate}
    
        \item {\bf Sequence similarity}. The similarity between the amino acids sequences of the proteins is usually considered as a relevant factor in determining whether the proteins could interact or not. However, this task should be performed cautiously, since, as stated by Luck et al. \cite{huri3}, \textit{proteins with higher sequence similarity usually have higher interaction profile similarities and tend to not interact, since their interaction interfaces is similar, as opposed to complementary}. 
        
        To compute the sequence similarity, the proteins' sequences, represented as strings, are \textbf{aligned}, i.e. arranged so that their most similar elements are juxtaposed, to identify regions of similarity that may be a consequence of functional, structural or evolutionary relationships between the sequences\footnote{\url{https://en.wikipedia.org/wiki/Sequence_alignment}}. An example of amino acids sequences alignment is shown in figure \ref{fig:alignment}.
        
        \begin{figure}
        \centering
        \includegraphics[width=7cm]{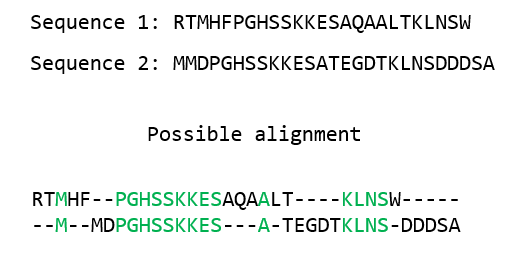}
        \caption{Example of amino acids sequences alignment. The most similar regions of the sequences are juxtaposed to identify possible similarities.}
        \label{fig:alignment}
        \end{figure}
        
        After two sequences are aligned, each comparison at each index \textit{i} could be either a \textit{match}, if the amino acid is the same, a \textit{mismatch}, if the amino acid is different, or a \textit{gap}, if a residue in one sequence is lacking in its counterpart. A visual representation of these three cases is presented in figure \ref{fig:match_mismatch_gaps}. 
        
        Since too many gaps could cause an alignment to become meaningless, gap penalties are usually added to adjust the similarity score based on the number and length of gaps.

        After the alignment, it is then possible to perform several operations (e.g. amino acids sostitutions) to measure the similarity between the sequences. One possible similarity measure is the Levenshtein distance, which is represented by the minimal edits to change one string into the other. However, this measure does not take into consideration that not all the amino acids substitutions are equal, since some amino acids are more similar than others. Indeed, we can divide the substitutions of two amino acids A and B into three categories:
        
        \begin{itemize}
        \item \textbf{Conservative}, if the characteristics of A and B are \textit{almost the same}.
        
        \item \textbf{Semi-conservative}, if A and B have \textit{similar} characteristics.
        
        \item \textbf{Non-conservative}, if A and B have \textit{different} characteristics.
        
        \end{itemize}
        
        Therefore, to compute the sequence similarity, usually more complex approaches, i.e. those which take into account the similarity of the amino acids characteristics, are used. 
        
        \begin{figure}
        \centering
        \includegraphics[width=7cm]{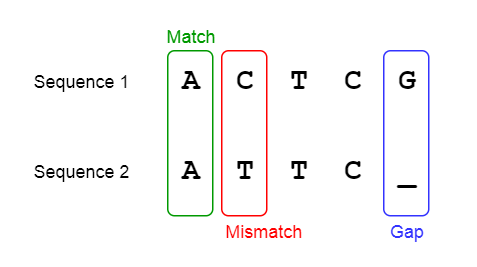}
        \caption{A match (in green) is when, in a specific position, the same amino acid is found in both sequences. Instead, when the amino acids are different is a mismatch (in red). Finally, a gap (in blue) is represented by the case in which an amino acid is not found in one of the two sequences\protect\footnotemark.}
        \label{fig:match_mismatch_gaps}
        \end{figure}
        
        \footnotetext{\url{https://towardsdatascience.com/pairwise-sequence-alignment-using-biopython-d1a9d0ba861f}}
        
        For each pair of proteins, both the global and the local sequences' similarities have been computed. The difference between these two similarities is that the former is computed starting from an \textbf{end-to-end} alignment of the entire sequences (as the one in figure \ref{fig:alignment}), while the latter is computed based on one or more alignments describing the most similar region(s) of the two sequences. 
        
        To accomplish this task, first the two input proteins' sequences were retrieved from Uniprot\footnote{https://www.uniprot.org/}, then they were aligned using the Bio.pairwise2 module\footnote{\url{https://biopython.org/DIST/docs/api/Bio.pairwise2-module.html}}. After the alignment, for each index i, a substitution matrix, BLOSUM 62,  was used to compute the similarity score of the amino acids at that index. The BLOSUM62 matrix has a dimension NxN, where N is the number of amino acids. A very high score in a cell $B_{IJ}$ means that the relative amino acids (the one in row I and the one in column J) are very likely to be exchanged, while a very low score means that their substitution is very unlikely. Therefore, from this matrix, a score is obtained from each pair of amino acids (those in the same position) in the aligned sequences and at the end the scores are summed and the final similarity score is obtained. \\ 
        However, the computational time needed to retrieve these two features is very high. Indeed, it took more than 10 hours to compute the similarity scores for about 55000 interactions.
        
        \item {\bf Uniprot Keywords (UKs)}. The Uniprot Keywords, a set of words used to describe the proteins in different biological fields, have been retrieved for each protein. For example, the keyword \textit{cytoplasm}, having as category \textit{cellular location}, is used to indicate that the protein is found in the cytoplasm. Instead, the keyword \textit{Acyltransferase}, having as category \textit{Molecular function}, is used to indicate that the protein is an \textit{enzyme catalyzing the transfer of acyl- (RCO-) groups}. 
        
        In total, there are 10 Uniprot categories: 
        
        \begin{itemize}
            \item {\textbf{Biological process}}: This category contains keywords assigned to proteins because they are involved in a particular biological process. Example: \textit{Abscisic acid biosynthesis}.
            
            \item {\textbf{Cellular component}}: This category contains keywords assigned to proteins because they are found in a specific cellular or extracellular component. Example: \textit{Centromere}.
            
            \item {\textbf{Coding sequence diversity}}: This category contains keywords assigned to proteins because their sequences can differ, due to differences in the coding sequences such as polymorphisms, RNA-editing, alternative splicing. Example: \textit{Chromosomal rearrangement}. 
            
            \item {\textbf{Developmental stage}}: This category contains keywords assigned to proteins because they are expressed specifically in a given developmental stage. Example: \textit{Early protein}.
            
            \item {\textbf{Disease}}: This category contains keywords assigned to proteins because they are involved in a specific disease. Example: \textit{Epilepsy}.
            
            \item {\textbf{Domain}}: This category contains keywords assigned to proteins because they have at least one specimen of a specific domain. Example: \textit{Homeobox}.
            
            \item {\textbf{Ligand}}: This category contains keywords assigned to proteins because they bind, are associated with, or whose activity is dependent of some molecule. Example: \textit{ATP-binding}.
            
            \item {\textbf{Molecular function}}: This category contains keywords assigned to proteins due to their particular molecular function. Example: \textit{Calmodulin-binding}.
            
            \item {\textbf{Post-translational modification}}: This category contains keywords assigned to proteins because their sequences can differ from the mere translation of their corresponding genes, due to some post-translational modification. Example: \textit{Acetylation}.
            
            \item {\textbf{Technical term}}: This category contains keywords assigned to proteins according to technical reasons. Example: \textit{Extinct organism protein}.
            
        \end{itemize}
        
        For each Uniprot category, an \textbf{overlap value} has been computed, denoting the number of keywords of that category present in both proteins. This value should incorporate the similarity of the two input proteins in that specific aspect. As revealed after an initial analysis, the category \textit{Technical term} turned out to be useless, and was removed. 
        
        The main weakness of the UKs is that sometimes they are missing for some categories of one or more of the interaction's proteins, making the comparison very difficult.
        
        \item \textbf{Gene Ontology Terms (GO terms)}. Like the Uniprot Keywords, the GO Terms are a \textit{controlled vocabulary to describe gene and gene product attributes in any organism}. However, unlike the UKs, the GO terms are developed independently of any existing database and can be classified only into one of the three Gene Ontology categories: biological process, molecular function and cellular component. 

        
        Again, for each interaction, three values, each of which denoting the number of terms shared by the two input proteins in a specific category, were obtained from these terms.
        
        \item \textbf{Complementary Domains}. As explained in section \ref{section:biological_background}, a protein domain is a \textit{a conserved region of a given protein sequence that can evolve, function, and exist independently of the rest of the protein chain}. The information about proteins domains are of crucial importance when trying to predict a protein-protein interaction, since only proteins with complementary domains can interact. Unfortunately, the knowledge about domain-domain interactions (i.e. complementary domains) is limited. 
        
        To retrieve the number of complementary domains for two input proteins, the proteins' domains were retrieved from Uniprot and the domain-domain interactions were retrieved both from Domine\footnote{\url{https://manticore.niehs.nih.gov/cgi-bin/Domine}} and 3did\footnote{\url{https://3did.irbbarcelona.org/}}. Then, for each interaction, the number of complementary domains between the two proteins was computed.

    \end{enumerate}
    
    It should be observed that no topological features were used in the model, although there are several methods than rely only on this type of features to perform protein-protein interactions prediction \cite{barabasippipred, rwppipred}. For example, Kovacs et al. \cite{barabasippipred} showed that two proteins connected by an high number of length three paths are most likely to be directly connected. However, the datasets used (especially the HQND) do not meet some necessary conditions that allow us to properly add the topological features without distorting the results. For example, there are proteins in the HQND dataset that are not in the training set. Therefore, the use of topological features would result in a penalty for the interactions containing these proteins (since the topological features of these interactions would be incorrect), and this would make the metrics biased. Furthermore, the dataset generated by the MFB-P method only contains negative interactions between proteins having the shortest path greater than 3, and this makes it unsuitable when topological properties are used. For example, a machine learning model using the "shortest path" feature, when trained on the above mentioned dataset, could learn, incorrectly, that each negative interaction would have a shortest path greater than 3 between its proteins. \\
    
    Finally, for each interaction, a vector of 15 values is created. The first two represent the global and the local similarities. The subsequent nine represent the number of keywords shared by the proteins in each of the nine Uniprot Categories\footnote{Biological process, cellular component, coding sequence diversity, developmental stage, disease, domain, ligand, molecular function, post-translational modification} (the category \textit{technical terms} was excluded). The subsequent three represent the number of GO terms shared by the proteins in each of the three Gene Ontology categories\footnote{biological process, molecular function, cellular component}, and the remaining value represents the number of complementary domains of the two proteins in the interaction.

\section{Models}
\label{section:models_used}

Also choosing the right model could significantly contribute to increasing the performance in the protein-protein interaction prediction task. Therefore, several classifiers were tested, sharing the set of features defined in the previous chapter, and their performances were compared. 

It should be noted that, given the type of task, the available data and the features used, probably a \textbf{deep learning} approach might not be the best solution. This is mainly due to two reasons:

\begin{itemize}
    \item The set of features is manually crafted. Therefore, we could not rely on the deep learning ability of automatically detect/create features.
    \item Especially in the medical field, the interpretability of the model (that is, basically, the knowledge about why a certain output is predicted and how the features contributed to this outcome) is of fundamental importance. Yet, one of the problem with deep learning is its poor interpretability.
\end{itemize}

The Random Forest Classifier, the Support Vector Machine and the Multi-layer Perceptron classifier were implemented using the Scikit-learn software\footnote{\url{https://scikit-learn.org/stable/}} \cite{scikit-learn}. Instead, the XGBoost library to perform eXtreme Gradient Boosting was downloaded from GitHub\footnote{\url{https://github.com/dmlc/xgboost}}. Finally, the Convolutional Neural Network was implemented using Keras\footnote{\url{https://keras.io/}}, an open-source library \cite{chollet2015keras}. 

Other than these five machine learning models, we also performed experiments with two of the sequence-based state-of-the-art protein-protein interactions prediction systems. Please note that it is not trivial to find the state-of-the-art models for PPIs link prediction, since often different datasets are used among models and also different training sets are adopted, making a fair comparison infeasible. Therefore, we picked the ones claiming to outperform the previous state-of-the art-models. The two chosen models are those presented in \cite{cnnlstm} and in \cite{CNNRandomModule}, and will be described in detail in subsections \ref{subsection:DNN-PPI} and \ref{subsection:DPPI}.

\subsection{Random Forest Classifier}

The Random Forest Classifier (RFC, \cite{randomforest}) is an ensemble model (i.e. a model that combines different machine learning algorithms) that creates a set of Decision Trees \cite{decisiontrees} from randomly selected subsets of the training set, and then combines the votes of these trees to decide the final result\footnote{\url{https://medium.com/machine-learning-101/chapter-5-random-forest-classifier-56dc7425c3e1}}. 

One of the main strengths of this model comes from the fact that the low correlation among the decision trees enables the RFC to avoid the individual errors of the models, following a more powerful "wisdom of crowds" principle. 

Some of the advantages of this model are: 

\begin{itemize}
    \item Since it is an ensamble of models, it has the ability to mitigate the overfitting issue.
    \item The input data must not be scaled.
    \item It is able to compute the features' importance.
\end{itemize}

\subsection{Support Vector Machine}

The Support Vector Machine (SVM, \cite{svm}) is a model that aims to identify the linear discriminant function with the maximum margin, where the margin is defined as \textit{the width that the boundary could be increased by before hitting a data point}. When the instances of a particular dataset are not linearly separable, usually the \textbf{kernel trick} is applied, which consists in mapping the points (i.e. instances) to a higher-dimensional space, in order to make them linearly separable. The above-mentioned kernel trick is probably one of the main strengths of this model, since, with an appropriate kernel, also complex problems can be solved.

\subsection{Multi-layer Perceptron classifier}

The Multi-layer Perceptron is an artificial neural network consisting of several layers: an input layer receiving the vector of features, an output layer that makes predictions about the input and one or more hidden layers, able to approximate any continuous function\footnote{\url{https://skymind.ai/wiki/multilayer-perceptron}}.

Generally, for each training iteration, two steps are performed:

\begin{enumerate}
    \item \textbf{Forward pass}. In this step, the signals go from the input layer to the output layer, passing through the hidden layer(s). Then, the output of the model is compared with the ground truth label, and the error is observed.
    \item \textbf{Backward pass}. In this step, the weights of the neural network are adjusted, using a backpropagation algorithm, i.e. an approach that uses the gradient descent method in order to find the minimum of the error function, following the \textit{chain rule\footnote{\url{https://en.wikipedia.org/wiki/Chain_rule}}}. 
\end{enumerate}

\subsection{Gradient Boosting}

The Gradient Boosting (GB) is a machine learning approach that creates a model by making an ensemble of weak prediction models, such as Decision Trees \cite{decisiontrees}. It was implemented through the XGBoost library.

Even though this approach might seem similar to the Random Forest approach when the decision trees are used, these methods have some important differences\footnote{\url{https://medium.com/@aravanshad/gradient-boosting-versus-random-forest-cfa3fa8f0d80}}:

\begin{itemize}
    \item One of the differences is the ensemble method used. Indeed, the Random Forest uses the bagging technique whereas the Gradient Boosting algorithm uses the boosting technique.
    \item The Gradient Boosting creates the trees one at the time, in a sequential way, increasing the execution time. Instead, the Random Forest approach trains each tree independently, using subsets of the training data. 
\end{itemize}

\subsection{Convolutional Neural Networks}

The convolutional neural networks (CNN, \cite{cnn}) are a class of deep neural networks. Although they are generally used in computer vision tasks, for their ability to detect and extract features from images, they are also widely used in the PPIS prediction task \cite{CNNRandomModule, cnnlstm} and in systems that aim to extract protein-protein interactions from the literature through text mining \cite{deeplearningforppi, multichannelCNN}. For example, the authors of \cite{CNNRandomModule} leverage the ability of the Convolutional Neural Network of \textit{learning a set of filters and detecting patterns in a protein sequence}. 

However, differently from the above mentioned systems, here the Convolutional Neural Networks are used with a restricted and hand-crafted set of features, which could limit their potential.

\subsection{DNN-PPI}
\label{subsection:DNN-PPI}

The deep neural network framework for predicting protein-protein interactions (DNN-PPI) was presented by Li et al. in \cite{cnnlstm}. 

The model works as follows:

\begin{enumerate}
    \item First, each amino acid in the protein sequence is encoded, i.e. replaced by a random natural number.
    \item Then, an embedding, i.e. a vectorial representation, is created for each amino acid, following the word2vec model \cite{word2vec}, considering the amino acids as words and the proteins' sequences as documents. 
    \item The proteins' sequences, represented by the concatenation of the embeddings of their amino acids, are given as input to three convolutional neural networks, each one followed by a max-pooling operation.
    \item Finally, the result of the last convolutional neural network is given as input to a Long Short-Term Memory (LSTM), responsible for learning the long-term relationships between the amino acids in the sequence. 
\end{enumerate}

This model claimed to obtain very high prediction accuracies (from 92.8\% to 97.9\%) on several external \textbf{positive} datasets: \textit{HPRD}\footnote{\url{http://www.hprd.org/}}, \textit{DIP}\footnote{\url{https://dip.doe-mbi.ucla.edu/dip/Main.cgi}}, \textit{HIPPIE}\footnote{\url{http://cbdm-01.zdv.uni-mainz.de/~mschaefer/hippie/}}, \textit{inWeb\_inbiomap}\footnote{\url{https://www.intomics.com/inbio/map.htmlsearch}}.

As negative interactions, the authors used interactions generated by the different subcellular location approach (section \ref{section:different_subcellular_location}).

\subsection{DPPI}
\label{subsection:DPPI}

The DPPI framework is another state-of-the-art system that was presented in \cite{CNNRandomModule}. In this model, the input proteins are first transformed into two $n \times 20$ matrices called \textbf{sequence profiles}, where $n$ is the length of the sequence. This representation is based on a large corpus of unsupervised data. Specifically, in these $n \times 20$ matrices, a cell $M_{i_j}$ represents the probability of $j_{t_h}$ amino acid in the $i_{t_h}$ position of the sequence. 

Then, these sequence profiles are given as input to several convolutional modules, each one composed of four layers: the convolutional layer, the activation layer, the batch normalization layer and the pooling layer. Then, the result of these modules is given as input to a Random Projection (RP) module, consisting of two separate fully-connected networks, that projects the inputs of the convolutional modules to two different spaces. The RP module is of key importance, since it enables the system to ignore the order of the input proteins and is able to detect relationships between the proteins.

\begin{figure}
\centering
\includegraphics[width=1\textwidth]{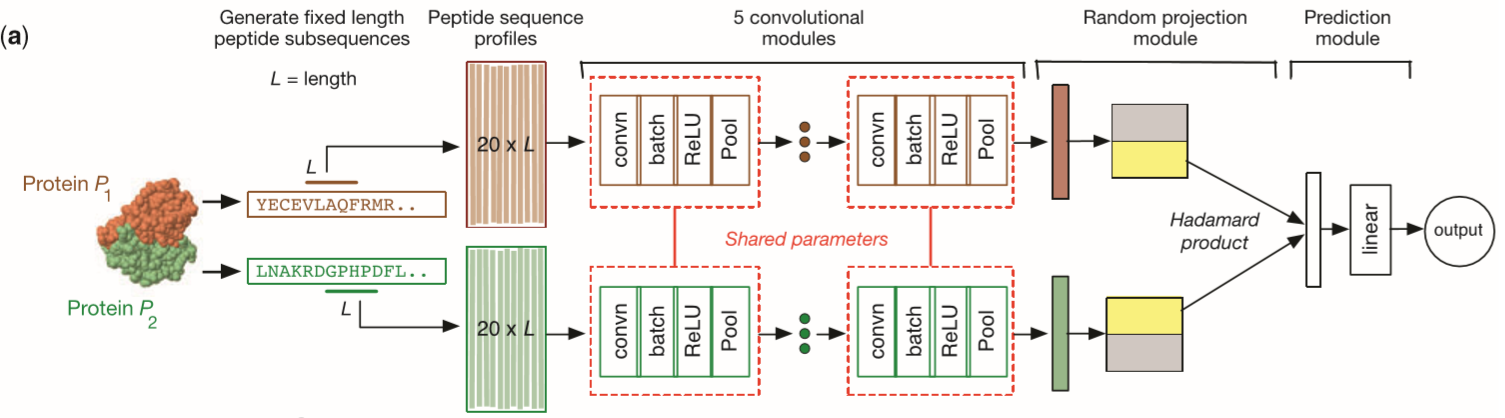}
\caption{The functioning of the DPPI prediction system \cite{CNNRandomModule}. The proteins' sequences are first transformed into sequence profiles, and then they are given as input to 5 convolutional modules. Later, the outputs of these modules are given as input to the random projection module, and the outputs of the latter are combined using the Hadamard product. At the end, a linear layer is used to compute the final output.}
\label{fig:dppi_system}
\end{figure}

Finally, The Hadamard product of the two outputs of the RP module (corresponding to the two representations of the input proteins) is performed, and the result is given as input to a linear layer to calculate the final result.
This method claims to outperform several state-of-the-art sequence-based PPI prediction methods, such as Profppikernel \cite{profppikernel} and PIPE2 \cite{pipe2}. A general overview of the system is represented in figure \ref{fig:dppi_system}.

Regarding the generation of negative interactions, this method adopted the random sampling strategy (section \ref{section:random_sampling}), generating ten negative interactions for each positive interaction. 

\section{Experiments}
\label{section:experiments_ppi_models}

For each of the machine learning models described in the previous section, several experiments were performed, comparing both the models themselves and how the accuracy of each specific model varied when different negative datasets were used in training.

\begin{table}[]
\centering
\begin{tabular}{|c|c|c|c|c|}
\hline
Training set & \begin{tabular}[c]{@{}c@{}}Train set\\ accuracy\end{tabular} & \begin{tabular}[c]{@{}c@{}}Test set\\ accuracy\end{tabular} & \begin{tabular}[c]{@{}c@{}}HQND\\ accuracy\end{tabular} & \begin{tabular}[c]{@{}c@{}}LIT-BM\\ accuracy\end{tabular} \\ \hline
\begin{tabular}[c]{@{}c@{}}HI-19 + \\ bait-prey\end{tabular} & 0.635 $\pm$ 0.001 & 0.636 $\pm$ 0.003 & 0.764 $\pm$ 0.015 & 0.708 $\pm$ 0.022\\ \hline
\begin{tabular}[c]{@{}c@{}}HI-19 + \\ MB-P\end{tabular} & 0.633 $\pm$ 0.001 & 0.631 $\pm$ 0.002 & 0.754 $\pm$ 0.01 & 0.7 $\pm$ 0.015 \\ \hline
\begin{tabular}[c]{@{}c@{}}HI-19 + \\ MFB-P\end{tabular} & \textbf{0.643 $\pm$ 0.002} & \textbf{0.644 $\pm$ 0.002} & 0.746 $\pm$ 0.001 & \textbf{0.737 $\pm$ 0.018} \\ \hline
\begin{tabular}[c]{@{}c@{}}HI-19 + balanced\\  random sampling\end{tabular} & 0.627 $\pm$ 0.002 & 0.626 $\pm$ 0.003 & \textbf{0.782 $\pm$ 0.02} & 0.7 $\pm$ 0.028 \\ \hline
\end{tabular}
\caption{Performances of the Random Forest Classifier when varying the negative training set used. Except for the accuracy on the HQND, the best results are obtained when the negative interactions generated by the MFB-P method are used for training.}
\label{table:RFC_performances}
\end{table}

\begin{table}[]
\centering
\begin{tabular}{|c|c|c|c|c|}
\hline
Training set & \begin{tabular}[c]{@{}c@{}}Train set\\ accuracy\end{tabular} & \begin{tabular}[c]{@{}c@{}}Test set\\ accuracy\end{tabular} & \begin{tabular}[c]{@{}c@{}}HQND\\ accuracy\end{tabular} & \begin{tabular}[c]{@{}c@{}}LIT-BM\\ accuracy\end{tabular} \\ \hline
\begin{tabular}[c]{@{}c@{}}HI-19 + \\ bait-prey\end{tabular} & 0.696 $\pm$ 0.002 & 0.623 $\pm$ 0.002 & 0.752 $\pm$ 0.01 & 0.58 $\pm$ 0.002\\ \hline
\begin{tabular}[c]{@{}c@{}}HI-19 + \\ MB-P\end{tabular} & 0.696 $\pm$ 0.0006 & 0.623 $\pm$ 0.0006 & 0.748 $\pm$ 0.008 & 0.578 $\pm$ 0.004 \\ \hline
\begin{tabular}[c]{@{}c@{}}HI-19 + \\ MFB-P\end{tabular} & \textbf{0.702 $\pm$ 0.001} & \textbf{0.63 $\pm$ 0.002} & 0.742 $\pm$ 0.006 & \textbf{0.59 $\pm$ 0.003} \\ \hline
\begin{tabular}[c]{@{}c@{}}HI-19 + balanced\\  random sampling\end{tabular} & 0.693 $\pm$ 0.001 & 0.617 $\pm$ 0.002 & \textbf{0.769 $\pm$ 0.006} & 0.589 $\pm$ 0.007 \\ \hline
\end{tabular}
\caption{Performances of the Support Vector Machine when varying the negative training set used. Except for the accuracy on the HQND, the best results are obtained when the negative interactions generated by the MFB-P method are used for training.}
\label{table:SVM_performances}
\end{table}

Table \ref{table:RFC_performances} summarizes the performance of the Random Forest Classifier. As we can see from the table, the RFC performed best when the negative interactions generated by the modified and filtered bait-prey method were used, except for the predictions of the negative interactions in the HQND validation set, where the negative interactions derived from the balanced random sampling led the model to perform better. 

The model was trained using the default scikit-learn parameters, but limiting the maximum depth of each tree to three.

The experiments concerning the Support Vector Machine are instead shown in Table \ref{table:SVM_performances}. The dynamics are exactly the same of the previous experiment. That is, the Support Vector Machine performed best on the training set, the test set and the LIT-BM dataset when the negative interactions generated by the MFB-P method were adopted. 
Instead, the accuracy on the HQND dataset is higher when the balanced random sampling is used to train the model. 

\begin{table}[]
\centering
\begin{tabular}{|c|c|c|c|c|}
\hline
Training set & \begin{tabular}[c]{@{}c@{}}Train set\\ accuracy\end{tabular} & \begin{tabular}[c]{@{}c@{}}Test set\\ accuracy\end{tabular} & \begin{tabular}[c]{@{}c@{}}HQND\\ accuracy\end{tabular} & \begin{tabular}[c]{@{}c@{}}LIT-BM\\ accuracy\end{tabular} \\ \hline
\begin{tabular}[c]{@{}c@{}}HI-19 + \\ bait-prey\end{tabular} & 0.65 $\pm$ 0.0008 & 0.647 $\pm$ 0.003 & 0.771 $\pm$ 0.004 & 0.582 $\pm$ 0.005\\ \hline
\begin{tabular}[c]{@{}c@{}}HI-19 + \\ MB-P\end{tabular} & 0.65 $\pm$ 0.0006 & 0.647 $\pm$ 0.002 & 0.77 $\pm$ 0.006 & 0.582 $\pm$ 0.005 \\ \hline
\begin{tabular}[c]{@{}c@{}}HI-19 + \\ MFB-P\end{tabular} & \textbf{0.661 $\pm$ 0.0006} & \textbf{0.658 $\pm$ 0.002} & 0.765 $\pm$ 0.004 & 0.6 $\pm$ 0.006 \\ \hline
\begin{tabular}[c]{@{}c@{}}HI-19 + balanced\\  random sampling\end{tabular} & 0.636 $\pm$ 0.0008 & 0.633 $\pm$ 0.003 & \textbf{0.805 $\pm$ 0.004} & \textbf{0.611 $\pm$ 0.008} \\ \hline
\end{tabular}
\caption{Performances of the Gradient Boosting when varying the negative training set used. The best results on the validation sets are achieved when using, for training, the negative interactions generated by the balanced random sampling method.}
\label{table:XGBOOST_performances}
\end{table}

Generally, the Random Forest Classifier performed better than the Support Vector Machine, especially on the LIT-BM dataset, underlying how the SVM was not able to learn the right features to identify positive interactions. The Support Vector Machine, for its part, achieves better accuracies on the training set. However, as pointed out by the performances on the external high-quality validation sets (HQND and LIT-BM), the SVM also seemed to learn the noise of the training set, overfitting.

Regarding the Gradient Boosting, the situation is slightly different. Indeed, as shown in Table \ref{table:XGBOOST_performances}, it obtained the highest accuracy on both the validation sets when the negative interactions generated from the balanced random sampling method were used for training. However, although the accuracy on the HQND tended to be high, the one on the LIT-BM dataset remained low, reaching a maximum value of only 0.61.

Also the Multilayer Perceptron, as shown in Table \ref{table:MLP_performances}, model achieved good accuracies on the HQND validation set but performed poorly on the LIT-BM validation set. Indeed, although the accuracy on negative interactions reached a peak of \textasciitilde 0.82, the accuracy on positive interactions remained very low, slightly better than random guessing.

\begin{table}[]
\centering
\begin{tabular}{|c|c|c|c|c|}
\hline
Training set & \begin{tabular}[c]{@{}c@{}}Train set\\ accuracy\end{tabular} & \begin{tabular}[c]{@{}c@{}}Test set\\ accuracy\end{tabular} & \begin{tabular}[c]{@{}c@{}}HQND\\ accuracy\end{tabular} & \begin{tabular}[c]{@{}c@{}}LIT-BM\\ accuracy\end{tabular} \\ \hline
\begin{tabular}[c]{@{}c@{}}HI-19 + \\ bait-prey\end{tabular} & 0.637 $\pm$ 0.001 & 0.637 $\pm$ 0.003 & 0.768 $\pm$ 0.012 & 0.51 $\pm$ 0.04\\ \hline
\begin{tabular}[c]{@{}c@{}}HI-19 + \\ MB-P\end{tabular} & 0.636 $\pm$ 0.002 & 0.636 $\pm$ 0.003 & 0.771 $\pm$ 0.013 & 0.494 $\pm$ 0.04 \\ \hline
\begin{tabular}[c]{@{}c@{}}HI-19 + \\ MFB-P\end{tabular} & \textbf{0.646 $\pm$ 0.008} & \textbf{0.646 $\pm$ 0.009} & 0.758 $\pm$ 0.02 & \textbf{0.52 $\pm$ 0.04} \\ \hline
\begin{tabular}[c]{@{}c@{}}HI-19 + balanced\\  random sampling\end{tabular} & 0.628 $\pm$ 0.001 & 0.627 $\pm$ 0.002 & \textbf{0.817 $\pm$ 0.007} & 0.504 $\pm$ 0.04 \\ \hline
\end{tabular}
\caption{Performances of the Multilayer Perceptron when varying the negative training set used. Except for the accuracy on the HQND, the best results are obtained when the negative interactions generated by the MFB-P method are used for training.}
\label{table:MLP_performances}
\end{table}

Regarding the Convolutional Neural Network, the results were very biased. In fact, it did not learn how to separate positive examples from negative ones, reaching an accuracy of 0.503 on both the training and the test sets. Regarding the validation sets, the CNN achieved the perfect accuracy on the LIT-BM set, while it achieved 0 accuracy on the HQND set, showing how the model was very inclined to predict an interaction as positive.

As mentioned previously, also two state-of-the-art models, DPPI and DNN-PPI were tested.

The performance of the DNN-PPI system were found to be very low, especially on the LIT-BM validation set, as shown in table \ref{table:DNNPPI_performances}. Even if,  when specific negative interactions were used (e.g. those generated by the MFB-P method), the accuracy on the training and test sets were very high, the system tended to perform poorly on both validation sets. 

\begin{table}[]
\centering
\begin{tabular}{|c|c|c|c|c|}
\hline
Training set & \begin{tabular}[c]{@{}c@{}}Train set\\ accuracy\end{tabular} & \begin{tabular}[c]{@{}c@{}}Test set\\ accuracy\end{tabular} & \begin{tabular}[c]{@{}c@{}}HQND\\ accuracy\end{tabular} & \begin{tabular}[c]{@{}c@{}}LIT-BM\\ accuracy\end{tabular} \\ \hline
\begin{tabular}[c]{@{}c@{}}HI-19 + \\ bait-prey\end{tabular} & 0.57 & 0.571 & 0.32 & 0.528\\ \hline
\begin{tabular}[c]{@{}c@{}}HI-19 + \\ MB-P\end{tabular} & 0.89 & 0.867 & 0.692 & 0.31 \\ \hline
\begin{tabular}[c]{@{}c@{}}HI-19 + \\ MFB-P\end{tabular} & 0.91 & 0.889 & \textbf{0.721} & 0.322 \\ \hline
\begin{tabular}[c]{@{}c@{}}HI-19 + balanced\\  random sampling\end{tabular} & 0.515 & 0.51 & 0.28 & 0.466 \\ \hline
\begin{tabular}[c]{@{}c@{}}Training set\\  used in \cite{cnnlstm}\end{tabular} & \textbf{0.951} & \textbf{0.951} & 0.071 & \textbf{0.941} \\ \hline
\end{tabular}
\caption{Performances of the DNN-PPI model when varying the negative training set used. Generally, the performances are very low. When their dataset is used, the model becomes biased and inclined to predict a new interaction as positive, as shown by the high accuracy on the positive validation set (LIT-BM) and the low accuracy on the negative validation set (HQND).}
\label{table:DNNPPI_performances}
\end{table}

Furthermore, when the training set used in the DNN-PPI paper, in which the negative interactions were generated by the different subcellular location approach, was adopted, the model performed very well on train and test sets and on the \textbf{positive} validation set (as also reported by them), while it performed very poorly on the negative validation set, confirming the hypothesis that the use of negative interactions generated by different subcellular locations makes the model inclined towards predicting a new interaction as positive.

Instead, the DPPI model performed better, as shown in table \ref{table:DPPI_performances}. Indeed, it generally obtained good accuracy on both training and test sets. However, while the performances on the LIT-BM validation set are very high, the one on the HQND validaiton set are just as good as simple random guessing. 

For the way the DPPI source code works, the model was trained on the entire training set. \\

As previously described, the models were evaluated on two validation sets, LIT-BM and HQND, analyzing how they performed in predicting, respectively, positive and negative protein-protein interactions. 

\begin{table}[]
\centering
\begin{tabular}{|c|c|c|c|c|}
\hline
Training set & \begin{tabular}[c]{@{}c@{}}Train set\\ accuracy\end{tabular} & \begin{tabular}[c]{@{}c@{}}Test set\\ accuracy\end{tabular} & \begin{tabular}[c]{@{}c@{}}HQND\\ accuracy\end{tabular} & \begin{tabular}[c]{@{}c@{}}LIT-BM\\ accuracy\end{tabular} \\ \hline
\begin{tabular}[c]{@{}c@{}}HI-19 + \\ bait-prey\end{tabular} & 0.76 & // & \textbf{0.502} & \textbf{0.99}\\ \hline
\begin{tabular}[c]{@{}c@{}}HI-19 + \\ MB-P\end{tabular} & \textbf{0.8} & // & 0.5 & \textbf{0.99} \\ \hline
\begin{tabular}[c]{@{}c@{}}HI-19 + \\ MFB-P\end{tabular} & 0.79 & // & 0.5 & \textbf{0.99} \\ \hline
\begin{tabular}[c]{@{}c@{}}HI-19 + balanced\\  random sampling\end{tabular} & 0.75 & // & 0.49 & \textbf{0.99} \\ \hline
\end{tabular}
\caption{Performances of the DPPI model when varying the negative training set used. The performances are very high on the LIT-BM validation set while they are low on the HQND validaiton set.}
\label{table:DPPI_performances}
\end{table}

However, a measure of the total goodness of a model, which is surely correlated with the accuracy of the model on the validation sets, is missing. The simple use of the arithmetic mean of the HQND accuracy and the LIT-BM accuracy would not allow us to distinguish between biased and non biased models. For example, a model with 1 accuracy on LIT-BM and 0.2 accuracy on HQND would have the same arithmetic mean of a model with 0.6 accuracy on both datasets, although the latter would be surely preferable since less biased.

Therefore, we decided to use the \textbf{harmonic mean} of the HQND accuracy and the LIT-BM accuracy, since, differently from the arithmetic mean, it penalizes very low performances on one of the two datasets. Formally, the harmonic mean of two variables $\alpha$ and $\beta$ can be defined as: \\

\begin{equation}
    2\frac{\alpha \beta }{\alpha + \beta}
\end{equation} \\

The overall comparison of all the models is presented in table \ref{table:models_comparison}. As shown in the table, the Random Forest Classifier is the best performing model, even when compared with the two state-of-the-art models. The DPPI system, although it performs very well when predicting positive interactions, has low results on the HQND datasets, and this leads it to a lower harmonic mean with respect to the RFC.

\begin{table}[]
\centering
\begin{tabular}{|c|c|c|c|c|c|}
\hline
\begin{tabular}[c]{@{}c@{}}Model\\ used\end{tabular} & \begin{tabular}[c]{@{}c@{}}Train set\\ accuracy\end{tabular} & \begin{tabular}[c]{@{}c@{}}Test set\\ accuracy\end{tabular} & \begin{tabular}[c]{@{}c@{}}HQND\\ accuracy\end{tabular} & \begin{tabular}[c]{@{}c@{}}LIT-BM\\ accuracy\end{tabular} &
\begin{tabular}[c]{@{}c@{}}Harmonic\\ mean\end{tabular} \\ \hline
RFC & 0.643 & 0.644 & 0.746 & 0.737 & \textbf{0.741}\\ \hline
SVM & 0.702 & 0.63 & 0.742 & 0.59 & 0.657\\ \hline
XGBoost & 0.636 & 0.633 & 0.805 & 0.611 & 0.695\\ \hline
MLP & 0.646 & 0.646 & 0.758 & 0.52 & 0.617\\ \hline
CNN & 0.51  & 0.5   & 0  &  1 & 0\\ \hline
DNN-PPI & 0.91 & 0.889 & 0.721 & 0.322 & 0.445\\ \hline
DPPI & 0.8 & // & 0.5 & 0.99 & 0.66\\ \hline
\end{tabular}
\caption{The overall comparison of the models. The Random Forest Classifier is the model that performs best. The DPPI is highly penalized from the low accuracy on the HQND validation set, which leads its harmonic mean to be lower than that of the RFC.}
\label{table:models_comparison}
\end{table}

\section{Features' importance}

\begin{figure}[]
    \centering
    \includegraphics[width=1\textwidth]{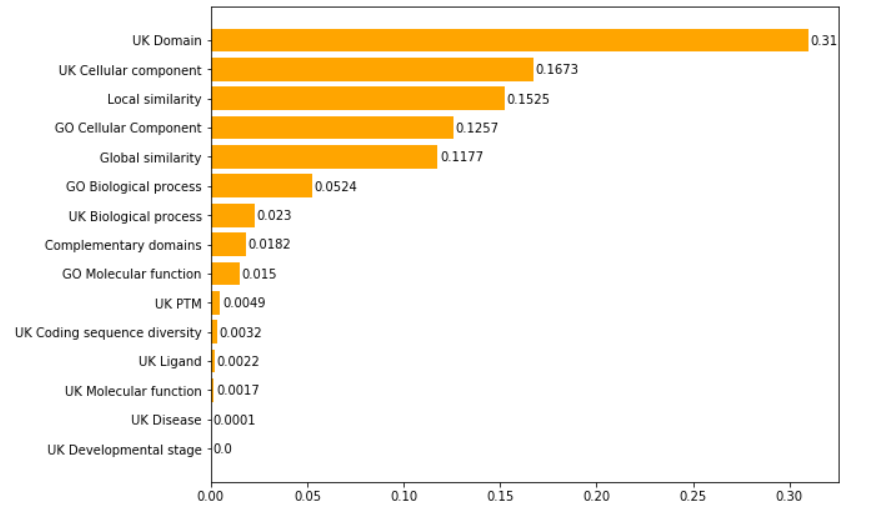}
    \caption{Importance of the features in the Random Forest Classifier. The most informative feature is the number of common domains, whereas the less informative is the common terms in the \textit{Developmental stage} category of the Uniprot Keywords.}
    \label{fig:feature_importances}
\end{figure}

As previously said, one of the advantages of the Random Forest Classifier is that it allows us to compute an estimation of the features' importance, that is shown in Figure \ref{fig:feature_importances}.

The number of common domains of the proteins is the most important feature by far (0.31), followed by the number of common cellular components (0.16) and the local similarity between the proteins' sequences (0.15). We also discovered that the number of complementary domains is particularly informative for the prediction of positive interactions, increasing the accuracy of the models by \textasciitilde 2\%. Surprisingly, the number of common diseases turned out to be almost irrelevant, also because of the fact that sometimes the Uniprot Keywords are missing for a specific protein. Moreover, the common terms in the category \textit{Developmental stage} turned out to be completely unimportant.

\chapter{PPIs extraction from biomedical literature}
\label{chapter:textmining_survey}

Thanks to the advances in the biomedical field, the size of the biomedical literature is rapidly growing\footnote{\url{https://www.nlm.nih.gov/bsd/medline_pubmed_production_stats.html}}. One of the most important sources of biomedical literature is MEDLINE\footnote{\url{https://en.wikipedia.org/wiki/MEDLINE}},  a bibliographic database of life sciences and biomedical information accessible through PubMed\footnote{\url{https://www.ncbi.nlm.nih.gov/pubmed/}}, which is a database that currently contains over 29.6 millions citations, of which over 1.3 millions were added in 2018\footnote{\url{https://www.ncbi.nlm.nih.gov/pubmed/?term=2018\%3A2018\%5Bdp\%5D}}. However, since the majority of the articles in PubMed are subject to traditional copyright restrictions, of major importance for text mining is the PubMed Open Access Subset\footnote{\url{https://www.ncbi.nlm.nih.gov/pmc/tools/openftlist/}} (PMOAS), a weekly updated subset of PubMed containing over 2.4 millions full-text freely available articles. 

The biomedical literature is particurarly challenging to text mining algorithms for several reasons \cite{textminingbiomed}. The first reason is that the writing style is different from the other types of literature, since it is {\bf more formal and complex}. Another important reason is that, when referring to biomedical processes and entities (e.g. genes, procedures, species), different terms can be used, and each one of them can have different spellings, abbreviations and database identifiers. This issue is usually tackled by applying {\bf entity recognition and normalization}. 

Several challenges have also been organized to underline the importance of  text mining in the biomedical field \cite{biocreativegeneral, challenges15years, overviewbionlp09}. One of the most important challenges is BioCreative. In the second and third editions of this challenge (BioCreative II and III) there was a task specific to protein-protein interaction extraction \cite{biocreative2overview} with the aim of:

\begin{itemize}
    \item \textit{exploring which approaches are successful and practical.}
    \item \textit{providing useful resources for training and testing protein interaction extraction systems.}
    \item \textit{analyzing the main difficulties and aspects influencing performance of PPI extraction systems.}
\end{itemize}

In addition to challenges specific to the biomedical domain, other challenges began to include tasks relative to the biomedical field. For example, SemEval, a series of semantic analysis evaluation organized each year, has included at least one task relevant to bioinformatics in the most recent editions. 

\section{Negative protein interactions detection}

By manually annotating a corpus of over 20000 sentences contained in biomedical research articles, Vincze et al. \cite{bioscope2008} reported that about 13\% of them contained one (or more) negation. More recently, Nawaz et al. \cite{negatedeventsanalysis}, analyzing three different biomedical corpus, reported that over 6\% of the bio-events are negated. Furthermore, given the relevance of these negated bio-events, the Journal of Negative Results in Biomedicine\footnote{\url{https://jnrbm.biomedcentral.com/}} has been launched in 2002 \cite{jnrbm}, with the aim of {\em encourage the publication of null results, addressing bias in the literature}. At the end of 2017, this journal ceased to be published, since, according to the authors,  {\em has succeeded in its mission and there is no longer a need for a specific journal to host these null results}. 

Despite the presence of the negative interactions in the biomedical literature, there are few methods that try to extract negative interactions using text mining algorithms. One of these methods has led to the creation of the Negatome 2.0, a dataset described in subsection \ref{subsection:negatome}, which however presents very low precision and recall.

Because of its complexity, the negative protein-protein interactions detection task can be divided in several sub-tasks:

\begin{enumerate}
    \item {\bf Gene entity recognition and normalization}: \\
    The aim of this task is to find all the occurrences of genes in the text (recognition) and to match each occurrence to an identifier belonging to a knowledge base that unequivocally represents its concept (normalization). For example, a protein could be referenced with its full name or with an acronym. The normalization algorithm should assign the same identifier to both occurrences.
    
    Previous studies suggest that this step is crucial to achieve good performances. For example, Sanchez et al. \cite{neganalysisandextraction} reported that the F1-score increased by 14\% after the normalization process was performed manually, ensuring the highest possible precision.
    
    The importance of this step is also highlighted by the creators of the BioCreative task about protein interaction extraction \cite{biocreative2overview}, saying that {\em common characteristic of the top scoring teams was the use of rather sophisticated interactor protein normalization strategies when compared with other systems} and that {\em is clear that using sophisticated gene mention and normalization detection strategies generally improved the results of participating teams and constitute one of the most important components for interaction extraction systems}. 
    
    The methods that have been developed to address this task can be divided into two categories: {\em dictionary-based approaches} and {\em machine learning} approaches. The first category of approaches generally creates a dictionary using information from online databases (e.g. UniProt) and then they perform string matching when examining a text. Instead, in the machine learning approaches, this subtask is considered as a sequence labelling task and annotated corpora are used to train the systems.
    
    \item {\bf Interaction articles detection (IAD)}: \\
    The aim of this task is to determine which articles are relevant to protein-protein interactions.
    
    This step is usually performed to reduce the number of articles that should be analyzed by the system, since usually only a small fraction of the articles contained in a corpus (e.g. PubMed) are relevant to PPIs. This subtask was also present in the BioCreative II and III tasks about protein-protein interaction \cite{biocreative2overview, biocreative3ppitask}, during which it was also released a manually annotated corpora of more than 15000 articles. 
    Altough the BioCreative II challenge is old (2006), the results of the IAD subtask were already promising, with the top team achieving a F1-score of 0.77, with a recall of 0.86 and a precision of 0.70. However, in a subsequent edition of the competition (BioCreative III, 2010), using different datasets, the top scoring team achieved a F1-score of just 0.63 in this task. Nonetheless, considering why this subtask is performed, it could be enough to achieve a very high recall (greater than 95\%) and a good precision (greater than 60\%). 
    
    As reported by Krallinger et al. \cite{biocreative2overview}, the main causes of false positives in the BioCreative II challenge were: (i) articles containing abstracts related to protein interactions that, however, did not contain protein-protein interactions that are worth annotating in the full text and (ii) articles describing interaction relations, but not between proteins.
    
    \item {\bf Interaction pair detection}:\\
    The aim of this task is to extract from the text the pairs of proteins that are considered to interact. Two annotated datasets are generally used for the training and the evaluation of this task: BioInfer \cite{bioinfer} and Aimed \cite{aimed}. 
    
    The approaches that attempt to tackle this problem can be divided into 3 categories:
    \begin{enumerate}
        \item The first category is represented by the methods based on {\bf statistics} that can be extrapolated from the entire corpus. For example, a pair of proteins could be considered as interacting based on co-occurrence analysis.
        \item The second category is represented by the {\bf rule-based} methods, i.e. methods that create a set of rules to extract the interactions. However, due to the size of the biomedical literature and to the complexity of the writing style, these approaches have limited usability.
        \item In the third category there are the {\bf machine learning} algorithms, mostly composed by {\em kernel methods}. As well explained by Tikk et al. \cite{kernelcomparison2013, kernelcomparison2010}, the kernel methods work by transforming each sentence into a structured representation that aims to best capture properties of how the interaction is expressed. These representations and their gold-standard labels are then used to train a kernel-based learner (e.g. SVM). 
        
        The state of the art among these kernel methods, expanding a convolution parse tree kernel with tree pruning methods and a decay factor \cite{singlekernelSOTA}, achieved 67\% F1-score on the Aimed dataset and 72.6\% F1-score on the BioInfer dataset. However, Tikk et al. \cite{kernelcomparison2010} showed, comparing several kernel methods in different evaluation settings, that {\em even the best performing kernels, requiring extensive parameter optimizations and large training corpora, cannot be considered as significantly better than a simple rule-based method which does not need any training at all and has essentially no parameters to tune}.
        
        Recently, to overcome the limitations of the kernel methods, approaches based on deep learning have been proposed \cite{rnnlstm, deeplearningforppi, multichannelCNN, ppiextractiondeep}. The majority of these systems use pre-trained word-embeddings (also in the biomedical domain \cite{distributionalresources}), sometimes adding additional features like the part of speech tags of the words. Then, as deep neural network model, often a CNN followed by a fully-connected layer is used. For example, Quan et al. \cite{multichannelCNN} presented a multichannel convolutional neural network. The main idea behind this method is to use, for each word, several embeddings representing the word in different context (i.e. embeddings trained on PubMed, embeddings trained on Wikipedia, etc). Applying this method, they achieved 72.4\% F1-score on Aimed and 79.6\% F1-score on BioInfer.
        
        More recently, Hsieh et al. \cite{rnnlstm} proposed a bidirectional LSTM to extract protein-protein interactions from the literature, relying on the recurrent neural network ability to better capture the long-term relationship among words. Their system achieved 76.9\% F1-score on Aimed and 87.2\% F1-score on BioInfer.
        \end{enumerate}

    \item {\bf Negation detection}:\\
    Negation detection has been identified as a main challenge in biomedical relation extraction \cite{negationimportant, negationimportant2}. Part of the complexity in identifying negation comes from the fact that it can be expressed in several ways. For example, it could be incorporated in the event-trigger, i.e. the word indicating the occurrence of the event, like the verb {\em dysregulate}, or it could be expressed even with the absence of an explicit negation cue\footnote{For example: Although MKK3, MKK4, and MKK6 all activated p38 MAPk in experimental models, {\bf only MKK3 was found to activate recombinant p38 MAPk in LPS-trated neutrophils}.}.
    
    Nawaz et al. \cite{negatedeventsanalysis} analyzed the main types of negated bio-events, reporting that the most common type is represented by the {\em negated triggers} (i.e. when an explicit negation, like not, modifies the event-trigger\footnote{ProteinA does {\bf not} interact with proteinB}), that account for approximately 60\% of the negated bio-events.
    
    Although this subtask is usually compared with the negation cue and scope identification task, i.e. the identification of the negation cue and of the words affected by it (scope), they are different. Indeed, a sentence might contain a negated bio-event without containing any negation cue, or, at the other end, it might contain a negation cue without containing any negated bio-events. This remarkable difference is also confirmed by Vincze et al. \cite{linguisticdifference}, after a comparison between an annotated corpus of negation scopes and a biologically annotated corpus of negated bio-events.
    In fact, they reported that: i) only 41\% of the bio-events containing an event-trigger inside the scope of a negation cue were actually negated and ii) 16\% of the negated bio-events had the event-triggers outside the scope of the negation cue. 
    
    However, a negation cue is still considered as {\em the most important factor to be considered} \cite{negatedeventsanalysis}. Consequently, the majority of the systems performing this task use a list of negation cues to extract the main features from a sentence. For example, Sarafraz et al. \cite{negationwithSVMcommand} used features such as whether a negation cue was present, the part of speech tag of the negation cue, the parse-tree distance between the event-trigger and the negation cue, etc. as input to a SVM, obtaining a F1-score of 51\%. More recently, Nawaz et al. \cite{negatedeventsanalysis} used a more complex negation cues list and more elaborate features (e.g. the presence, after the cue, of a deactivator, i.e. a word that deactivates the negation cue effect\footnote{e.g. {\em only} and {\em clear} are deactivators of the negation {\em not}}), achieving a F1-score of over 70\%.
    
    \item {\bf Interaction method extraction}:\\
    The aim of this task is to extract the method that has been used to characterize the interaction between a pair of proteins. This subtask is very important since the degree of reliability of the reported interaction is strongly correlated with the interaction method used to express it \cite{biocreative2overview, hi14}.
    
    \item {\bf Interaction sentence detection}:\\
    The aim of this task is to extract the sentence that best expresses the interaction found. The sentence retrieved by this task could help in the interpretation, update and evaluation of annotations.

    \end{enumerate}
    
\chapter{Conclusions}
\label{chapter:conclusions}

This thesis aimed to discuss and examine the issue of negative protein-protein interactions, demonstrating how this problem leads to the overestimation of the performances of some PPIs prediction systems. 

First, we described the methods for generating NPIs, showing that each method has its own weaknesses, although some of them are still widely used. Then, we compared the less biased methods using highly reliable training and test sets, concluding that the Modified and Filtered Bait-Prey approach can be considered the most effective. However, this method, given the shortest path heuristic on which it is based, could lead to the creation of biased datasets, and in this case the NPIs generated from the balanced random sampling should be used as training negative instances, as highlighted in chapter \ref{chapter:ppi_prediction_model}. Our experiments also showed that the approximated true error rates of the methods for generating negative instances are very low, revealing that the method should be picked mainly according to the bias of the dataset generated by it.

We also showed, using extremely reliable datasets, how the performances of two state-of-the-art PPIs prediction methods are overestimated, and we developed a PPIs prediction system which achieved better results with respect to these two methods, demonstrating that good performances are still achievable in the PPIs prediction task.


\section{Future work}

Exactly as the positive interactions, also the negatives could be extracted from the biomedical literature, since it is shown that over 6\% of bio-events are negated. An attempt was made by the creators of Negatome 2.0, which, however, presents very low precision and recall, mainly due to the simplicity of the text mining techniques used. Hence, even if the extraction of NPIs has not been a contribution of this work, we provided a survey on the extraction of negative protein-protein interactions from the literature, which can be taken as a reference point from future works.

Finally, machine learning specialist should collaborate with biological domain experts in order to develop a new PPIs prediction system which can finally overcome the limitations of the high-throughput experiments generally used to discover new protein-protein interactions.

\clearpage
\bibliographystyle{plain}
\bibliography{references}
\end{document}